\documentclass[twocolumn,showpacs,amsmath,amssymb,superscriptaddress]{revtex4-1}
\usepackage{dcolumn}
\usepackage{color}
\usepackage{graphicx}
\usepackage{amsmath}
\usepackage{multirow}
\usepackage{bm}
\usepackage{url}

\newcommand\pt{(p,t)}
\newcommand\tp{(t,p)}
\newcommand\mbos{$m$-IBM{\,}}
\newcommand\pbos{$p$-IBM{\,}}

\begin{document}

\title{
Two-neutron transfer reactions and shape phase transitions in the microscopically-formulated interacting
boson model
}

\author{K. Nomura}
\affiliation{Physics Department, Faculty of Science, University of
Zagreb, HR-10000 Zagreb, Croatia}
\affiliation{Advanced Science Research Center, Japan Atomic Energy
Agency, Tokai, 319-1195 Ibaraki, Japan}

\author{Y. Zhang}
\affiliation{Department of Physics, Liaoning Normal University, Dalian 116029, People's Republic of China}

\date{\today}

\begin{abstract}
Two-neutron transfer reactions are studied 
 within the interacting boson model based on the nuclear energy density
 functional theory. 
Constrained self-consistent mean-field calculations with the Skyrme
 energy density functional are performed to provide microscopic input to completely 
 determine the Hamiltonian of the IBM. 
Spectroscopic properties are calculated only from the nucleonic degrees of freedom. 
This method is applied to study the $\tp$ and $\pt$ transfer
 reactions in the assorted set of rare-earth
 nuclei $^{146-158}$Sm, $^{148-160}$Gd, and $^{150-162}$Dy, where
 spherical-to-axially-deformed shape phase transition is suggested to
 occur at the neutron number $N\approx 90$. 
The results are compared with those from the purely phenomenological IBM
 calculations, as well as with the available experimental data. 
The calculated $\tp$ and $\pt$ transfer reaction
 intensities, from both the microscopic and phenomenological IBM
 frameworks, signal the rapid nuclear structural change at particular
 nucleon numbers.
\end{abstract}


\maketitle


\section{Introduction}


The simultaneous theoretical description of nuclear structure and
reaction is one of the ultimate goals of low-energy nuclear physics. 
At experiment nucleon-pair transfer reactions are instrumental for
studying variety of nuclear structure phenomena. 
Of particular interest here is the shape phase transition  
\cite{cejnar2009,cejnar2010,iachello2011a,carr-book}, where nuclear
shape/structure changes as a function 
of nucleon number and which is identified as an abrupt change of
observables that are considered the order parameters of the phase transition. 
For many decades the two-nucleon transfer reactions, especially the $\tp$ and $\pt$ ones,
have been used  to study rapid structural evolution from 
one nuclear structure to another 
\cite{maxwell1966,BJERREGAARD1966145,fleming1971,casten1972,fleming1973,SHAHABUDDIN1980109,lovhoiden1986,LOVHOIDEN1989157,lesher2002} 
and, in that context, explored by a number of empirical theoretical
models \cite{fossion2007,clark2009,zhang2017,cejnar2010}.

The Interacting Boson Model (IBM) \cite{IBM}, a model where correlated nucleon
pairs are represented by bosonic degrees of freedom, has been
remarkably successful in the phenomenological description of low-energy
collective excitations in medium-heavy and heavy nuclei. 
The microscopic foundation of the IBM, starting from nucleonic degrees of
freedom, has been explored for decades 
\cite{OAIT,OAI,mizusaki1997,nomura2008,nomura2011rot}. 
Among these studies, a comprehensive method to derive the Hamiltonian of
the IBM has been developed in Ref.~\cite{nomura2008}. 
In this method, potential energy surface (PES) in the quadrupole deformation
space is calculated within the
constrained self-consistent mean-field (SCMF) method with a choice
of energy density functional (EDF), and is mapped onto the expectation value of the IBM
Hamiltonian in the boson coherent state \cite{ginocchio1980}. 
This procedure uniquely determines the strength parameters of the IBM
Hamiltonian. 
For strongly-deformed nuclei in particular, rotational response
of the nucleonic intrinsic state has been incorporated 
microscopically in the IBM framework, and this has allowed for
calculating the rotational spectra of deformed nuclei accurately \cite{nomura2011rot}. 
Since the EDF framework provides a global mean-field description of
various low-energy properties of the nuclei over the entire region of
the nuclear chart, it has become possible to derive the IBM Hamiltonian
for any arbitrary nuclei in a unified way.

In this article, we present a first application of the SCMF-to-IBM mapping
procedure of Refs.~\cite{nomura2008,nomura2011rot} to the nucleon-pair transfer
reactions as a signature of the shape phase transitions. 
We demonstrate how the method works for the description of the
transfer reactions, in the applications to the rare-earth
nuclei $^{146-158}$Sm, $^{148-160}$Gd, and $^{150-162}$Dy, which are an
excellent example of the spherical-to-axially-deformed shape phase
transition \cite{cejnar2010}. 
To the best of our knowledge, ever since its first application in 1977
\cite{arima1977trf}, the IBM has not been used as extensively to describe
nuclear reactions, including the two-nucleon
transfer reactions, which involve different nuclei, as the spectroscopy in a single nucleus. 
There are a few recent examples where the IBM was used in
phenomenological studies of $\tp$ and $\pt$ reactions
\cite{fossion2007,pascu2009,pascu2010,zhang2017}.

Already in Ref.~\cite{kotila2012}, key spectroscopic properties of the
above-mentioned Gd and Dy nuclei, i.e., energies and electromagnetic
transition rates, that signal the
first-order phase transition, were studied within the SCMF-to-IBM mapping
procedure using the Skyrme SkM* \cite{bartel1982} EDF and were compared
with the purely phenomenological IBM calculation. 
The main conclusion of that study was that 
the shape transition as a function of the neutron number $N$ occurred 
rather moderately in the microscopically-formulated IBM, as compared to
the phenomenological IBM calculation \cite{kotila2012}.

Here we have made somewhat a similar analysis to the one in
\cite{kotila2012}, that is, compared the $(p,t)$ and $(t,p)$ transfer
reaction intensities obtained from the SCMF-to-IBM 
mapping procedure with those from the phenomenological IBM calculation of Ref.~\cite{kotila2012}. 
In addition, we also compare our results with a more recent, extensive 
IBM study for the $\tp$ and $\pt$ transfer reactions in the same mass
region \cite{zhang2017}. 
In this way, we shall examine the robustness of the IBM framework on the
pair-transfer reactions and shape phase transitions. 

In Sec.~\ref{sec:theory} we describe the theoretical methods. The
calculated potential energy surfaces, excitation spectra, and
$(p,t)$ and $(t,p)$ transfer reaction intensities for the considered
nuclei are presented in Sec.~\ref{sec:results}, followed by a concise summary and
concluding remarks in Sec.~\ref{sec:summary}.

\section{Theoretical tools\label{sec:theory}}

Firstly we briefly describe the SCMF-to-IBM mapping procedure,
together with the two other phenomenological IBM calculations, which
have been employed in the present work. More detailed accounts of the
employed theoretical methods have been
already given in Refs.~\cite{nomura2010,nomura2011rot,kotila2012,zhang2017}, and
the reader is referred to that literature.

\subsection{SCMF-to-IBM mapping}

In the present analysis we used the neutron-proton
IBM (IBM-2), which distinguishes both
neutron and proton degrees of freedom \cite{OAI}. 
The IBM-2 is comprised of the neutron (proton) monopole $s_\nu$ ($s_\pi$) and
quadrupole $d_{\nu}$ ($d_{\pi}$) bosons, which represent, from a
microscopic point of view, the collective pairs of valence neutrons
(protons) with spin and parity $0^+$ and $2^+$, respectively \cite{OAI}. 
The number of neutron (proton) bosons, denoted by $N_\nu$ ($N_\pi$), is
equal to that of the neutron (proton) pairs. 
In this work the doubly-magic nucleus $^{132}$Sn has been taken as an
inert core. Hence $1\leq N_\nu\leq 7$, and $N_\pi=6$ (for $^{146-158}$Sm),
$N_\pi=7$ (for $^{148-160}$Gd), and $N_\pi=8$ (for $^{150-162}$Dy). 
For the IBM-2 Hamiltonian we employed the following form: 
\begin{eqnarray}
\label{eq:ham}
 \hat H = \epsilon(\hat n_{d_\nu} + \hat n_{d_\pi})+\kappa\hat
  Q_{\nu}\cdot\hat Q_{\pi} + \kappa'\hat L\cdot\hat L,
\end{eqnarray}
where $\hat n_{d_\rho}=d^\dagger_\rho\cdot\tilde d_{\rho}$ ($\rho=\nu,\pi$) is the
$d$-boson number operator, $\hat Q_\rho=d_\rho^\dagger s_\rho +
s_\rho^\dagger\tilde d_\rho + \chi_\rho(d^\dagger_\rho\times\tilde
d_\rho)^{(2)}$ is the quadrupole operator, and $\hat L=\hat L_\nu + \hat
L_\pi$ is the angular momentum operator with $\hat
L_{\rho}=\sqrt{10}(d_\rho^\dagger\times\tilde d_\rho)^{(1)}$. 
$\epsilon$, $\kappa$, $\chi_\nu$, $\chi_\pi$, and $\kappa'$ are the
parameters.

As the first step of determining the IBM-2 Hamiltonian, 
we carried out for each considered nucleus the constrained SCMF
calculation within the Hartree-Fock+BCS method \cite{ev8} based on the Skyrme SkM*
EDF \cite{bartel1982} to obtain PES with the quadrupole $(\beta,\gamma)$ shape degrees of freedom. 
The constraint is that of the mass quadrupole moment and, for the pairing
correlation, the density-dependent $\delta$-type pairing force has been
used with the strength of 1250 MeV fm$^3$.

The SCMF PES thus obtained has been mapped onto the expectation value of the IBM-2
Hamiltonian in the boson coherent state \cite{ginocchio1980}, and this
procedure completely determined the parameters $\epsilon$, $\kappa$,
$\chi_\nu$, and $\chi_\pi$ \cite{nomura2008,nomura2010}. 
Only the strength parameter $\kappa'$ for
the $\hat L\cdot\hat L$ term has been determined separately from the
other parameters, by adjusting the cranking moment of inertia in the
boson intrinsic state to the corresponding cranking moment of inertia
computed within the SCMF calculation at the equilibrium mean-field minimum
\cite{nomura2011rot}. 
No phenomenological adjustment of the parameters to experiment was made
in the whole procedure. 
We used the same values of the parameters as used in
Ref.~\cite{nomura2010} for the Sm isotopes and Ref.~\cite{kotila2012} for the Gd
and Dy isotopes.

Energy spectra and electromagnetic transition rates have been
obtained by the $m$-scheme diagonalization of the mapped IBM-2 Hamiltonian
\cite{nomura2012phd}, and the resulting wave functions have been used to
calculate the $\tp$ and $\pt$ transfer reaction intensities. 
In this work, as in \cite{zhang2017}, we only considered the $\tp$ and $\pt$
transfers of the monopole and quadrupole pairs of neutrons within each isotopic chain. 
The corresponding $\tp$ and $\pt$ transfer operators, 
denoted by $\hat P^{(L)}_{+}$ and $\hat P^{(L)}_{-}$ (with $L=0$ or
$2$), respectively, can be expressed as \cite{arima1977trf,IBM}: 
\begin{eqnarray}
\label{eq:opt0}
 &&\hat P^{(0)}_{+}=(\hat P^{(0)}_{-})^{(\dagger)}=t_{0}s^{\dagger}_\nu
  A(\Omega_\nu,N_\nu) \\
\label{eq:opt2}
 &&\hat P^{(2)}_{+}=(\hat P^{(2)}_{-})^{(\dagger)}=t_{2}d^{\dagger}_\nu
  A(\Omega_\nu,N_\nu),
\end{eqnarray}
The factor $A(\Omega_\nu,N_\nu)$ in Eqs.~(\ref{eq:opt0}) and (\ref{eq:opt2}) is given by
\begin{eqnarray}
\label{eq:fac}
 A(\Omega_\nu,N_\nu)=\sqrt{\Omega_\nu-N_\nu-\hat n_{d_\nu}},
\end{eqnarray}
with $\Omega_\nu$ the degeneracy of the neutron pairs in a given major
shell, i.e., $\Omega_\nu=(126-82)/2=22$ in the considered nuclei. 
For the sake of simplicity, the operator $\hat n_{d_\nu}$ in
Eq.~(\ref{eq:fac}) has been replaced with its expectation value in the
ground state of the initial nucleus, i.e., $\langle\hat n_{d_\nu}\rangle_{0^+_1}$ \cite{fossion2007}. 
$t_0$ and $t_2$ in the same equation are overall scale factors. 
The intensities of the $\tp$ and $\pt$ transfer reactions are given, respectively, as: 
\begin{eqnarray}
\label{eq:tp}
&&I^{\rm (tp)}(N,J_i\to N+2,J_f) \nonumber \\
&&=\frac{1}{2J_i+1}|\langle N+2,J_f||\hat P_{+}^{(L)}||N,J_i\rangle|^2
\end{eqnarray}
and 
\begin{eqnarray}
\label{eq:pt}
&&I^{\rm (pt)}(N+2,J_i\to N,J_f) \nonumber \\
&&
=\frac{1}{2J_i+1}|\langle N,J_f||\hat P_{-}^{(L)}||N+2,J_i\rangle|^2,
\end{eqnarray}
where the state $|N,J_{i,f}\rangle$ represents the IBM-2 wave
function for a nucleus with the neutron number $N$ and total
angular momentum $J_i$ for the initial or $J_f$ for the final states.
Here we considered the transfer reactions from the $0^+_1$ ground state
of the initial nucleus to the lowest three $0^+$ and $2^+$ states of the final nucleus. 

In what follows, the mapped IBM-2 framework, described in this section, is referred to as \mbos.

%
%

\begin{figure*}[htb!]
\begin{center}
\includegraphics[width=0.7\linewidth]{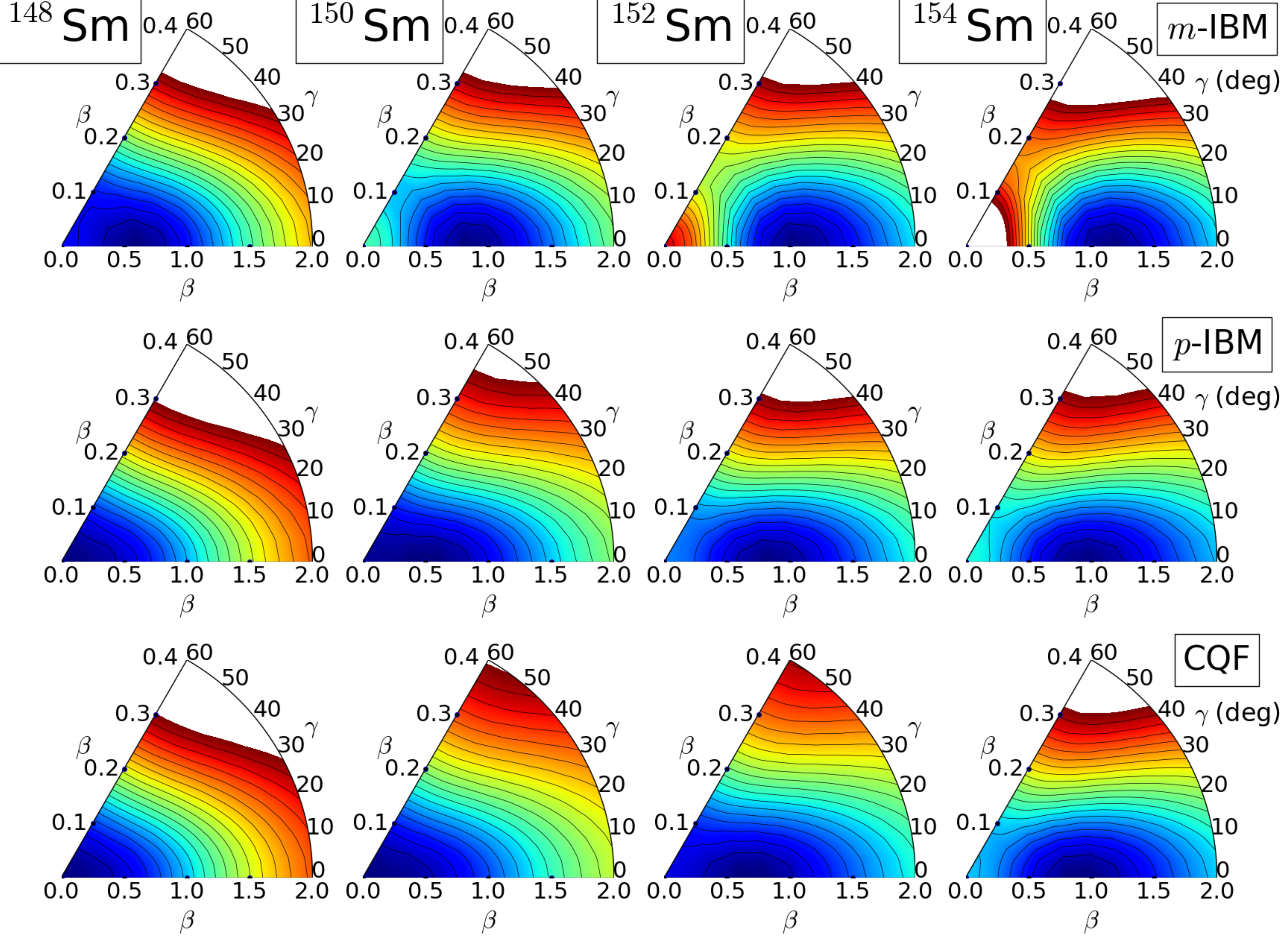} 
\caption{Potential energy surfaces for the nuclei $^{148-154}$Sm 
 plotted within the $(\beta,\gamma)$ deformation space and with up to 5
 MeV from the global minimum. The energy difference between neighboring
 contours is 250 keV. See the main text for details.}
\label{fig:pes_sm} 
\end{center}
\end{figure*}

\begin{figure*}[htb!]
\begin{center}
\includegraphics[width=0.7\linewidth]{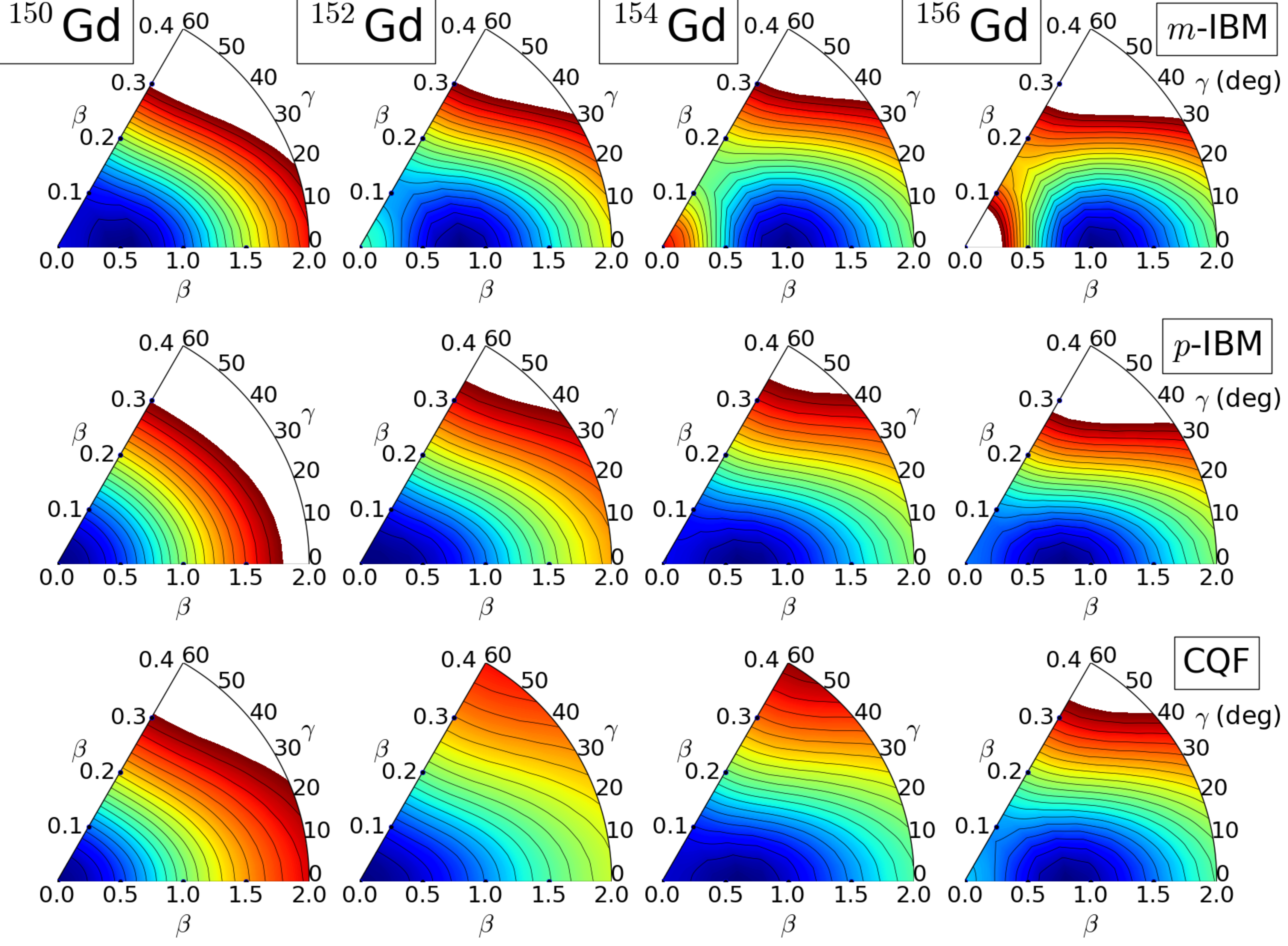} 
\caption{Same as for Fig.~\ref{fig:pes_sm}, but for the nuclei $^{150-156}$Gd.}
\label{fig:pes_gd} 
\end{center}
\end{figure*}

\begin{figure*}[htb!]
\begin{center}
\includegraphics[width=0.7\linewidth]{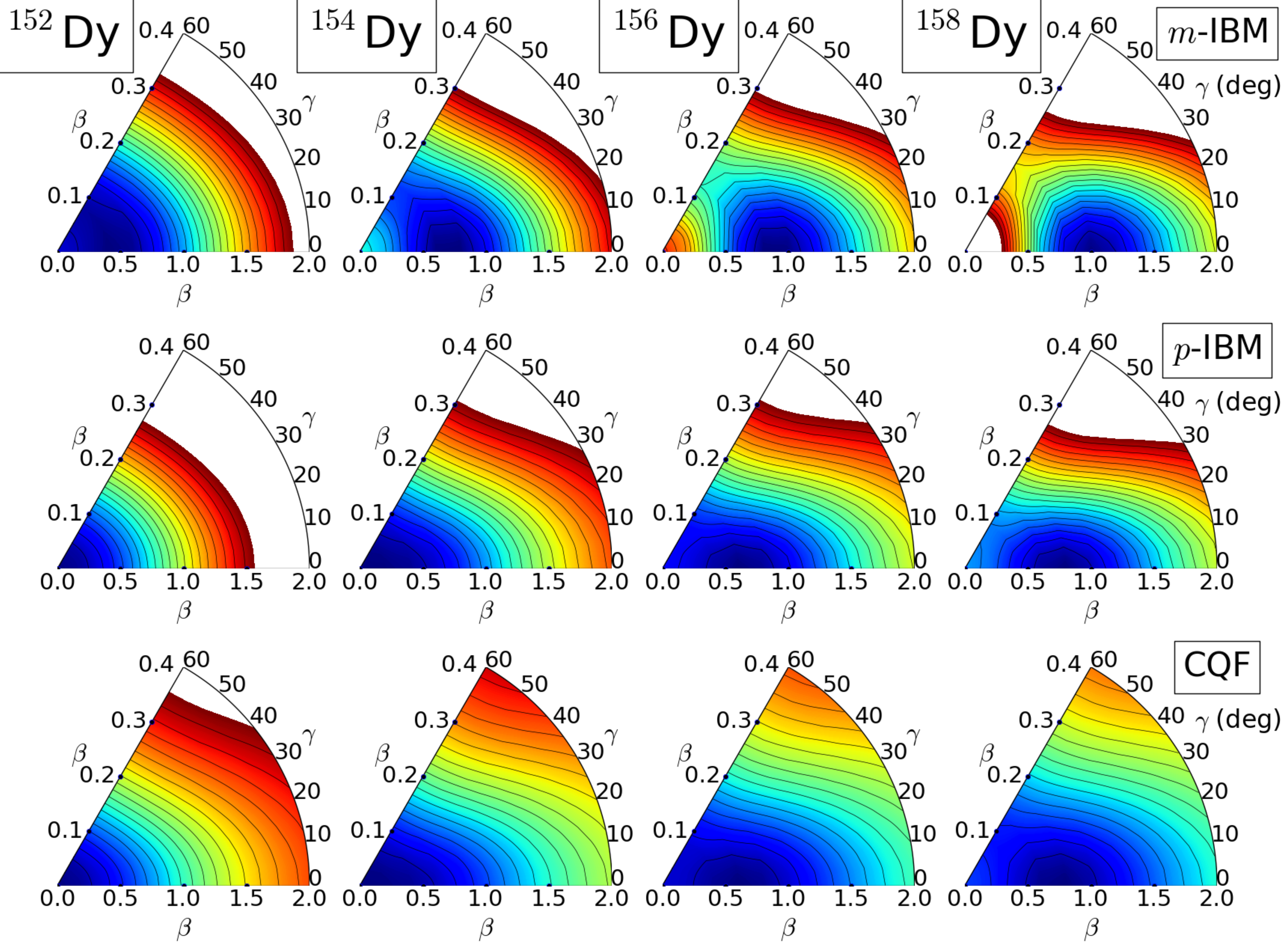} 
\caption{Same as for Fig.~\ref{fig:pes_sm}, but for the nuclei $^{152-158}$Dy.}
\label{fig:pes_dy} 
\end{center}
\end{figure*}

\subsection{Phenomenological IBM-2}

Along with the \mbos calculation we have carried out the purely phenomenological IBM-2
calculations using the same Hamiltonian as in Eq.~(\ref{eq:ham}), but with
parameters adjusted to reproduce low-energy spectra for each considered nucleus. 
The fitted parameters for the Sm isotopes are presented in
Table~\ref{tab:sm_para_ibm2}. 
The parameters for the nuclei $^{150}$Sm and $^{152}$Sm have been taken
from Ref.~\cite{scholten1980phd}. 
For the Gd and Dy isotopes, we employed the same values of the
parameters $\epsilon$, $\kappa$, $\chi_\nu$, and $\chi_\pi$ as those
used in Ref.~\cite{kotila2012}. 
The IBM-2 Hamiltonian considered in Ref.~\cite{kotila2012} was comprised, in addition
to the three terms in the above Hamiltonian in Eq.~(\ref{eq:ham}), those proportional to
$(d_\rho^\dagger\times d_\rho^{\dagger})^{(L)}\cdot (\tilde
d_\rho\times\tilde d_\rho)^{(L)}$ with $L=0$ and 2, and the so-called
Majorana terms. 
In the present calculation, these terms have not been included, as they
play only a minor role in the description of the low-lying states. 
The $(t,p)$ and $(p,t)$ transfer operators were already defined in 
Eqs.~(\ref{eq:opt0})--(\ref{eq:tp}). 

We denote, hereafter, the purely phenomenological
IBM-2 calculation thus far mentioned as \pbos, unless otherwise specified. 

\begin{table}
\caption{\label{tab:sm_para_ibm2} The \pbos parameters of the IBM-2
 Hamiltonian in Eq.~(\ref{eq:ham}) for the nuclei $^{146-158}$Sm,
 determined in this study so as to reproduce the experimental low-lying
 spectra. The value of the parameter $\kappa'$ has been taken to be zero
 for all the Sm nuclei.} 
\begin{center}
\begin{tabular}{cccccccc}
\hline\hline
& $^{146}$Sm & $^{148}$Sm & $^{150}$Sm & $^{152}$Sm & $^{154}$Sm &
 $^{156}$Sm & $^{158}$Sm \\
\hline
$\epsilon$ (MeV) & 1.100 & 1.000 & 0.700 & 0.520 & 0.450 & 0.400 &
			     0.400 \\
$\kappa$ (MeV) & -0.140 & -0.130 & -0.080 & -0.075 & -0.085 & -0.085 &
			     -0.085 \\
$\chi_\nu$ & -0.800 & -1.000 & -0.800 & -1.000 & -1.200 & -1.200 &
			     -1.200 \\
$\chi_\pi$ & -0.800 & -1.000 & -1.300 & -1.300 & -1.200 & -1.200 &
			     -1.200 \\
\hline\hline
\end{tabular} 
\end{center} 
\end{table}

\subsection{IBM-1 in the consistent-Q formalism}

We have also performed a similar phenomenological calculation within the IBM-1, where
no distinction is made between neutron and protons bosons. 
We adapted the same Hamiltonian in the so-called consistent-Q formalism (CQF) \cite{warner1983}
as the one used in Ref.~\cite{zhang2017}. 
The CQF Hamiltonian reads: 
\begin{equation}
\label{eq:ham-cqf}
 \hat H_\text{CQF}=\epsilon_0\Big[(1-\eta)\hat n_d -\frac{\eta}{4N}\hat Q^\chi\cdot\hat Q^\chi\Big].
\end{equation}
$\eta$ and $\chi$ (which appears in the quadrupole operator $\hat Q$)
are the control parameters, and $\epsilon_0$ is the scale factor fitted
to reproduce the $2^+_1$ excitation energy for each nucleus. 
The $(t,p)$ and $(t,p)$ transfer operators in the IBM-1 framework are
similar to the IBM-2 counterparts in Eqs.~(\ref{eq:opt0})--(\ref{eq:tp}),
except for the factor $A(\Omega_\nu,N_\nu)$. 
For all the details of the CQF calculation, the reader is referred to
Ref.~\cite{zhang2017}. 
For the calculations on the Sm and Gd isotopes, the same parameters 
as in Ref.~\cite{zhang2017} have been used. 
Only for the Dy isotopes, the calculation has been newly made, and the
values of the control parameters $\eta$ and $\chi$ for the Hamiltonian
$\hat H_\text{CQF}$ have been taken from the earlier IBM-1 study on the
rare-earth nuclei in Ref.~\cite{maccutchan2004} and are listed in
Table~\ref{tab:dy_para_cqf}. 

\begin{table}
\caption{\label{tab:dy_para_cqf} Control parameters of the Hamiltonian
 $\hat H_\text{CQF}$ in Eq.~(\ref{eq:ham-cqf}) determined in this study for the isotopes
 $^{150-162}$Dy.}
\begin{center}
\begin{tabular}{cccccccc}
\hline\hline
& $^{150}$Dy & $^{152}$Dy & $^{154}$Dy & $^{156}$Dy & $^{158}$Dy &
 $^{160}$Dy & $^{162}$Dy \\
\hline
$\eta$ & 0.1 & 0.35 & 0.49 & 0.62 & 0.71 & 0.81 & 0.92 \\
$\chi$ & -1.12 & -1.10 & -1.09 & -0.85 & -0.67 & -0.49 & -0.31 \\
\hline\hline
\end{tabular} 
\end{center} 
\end{table}

\section{Results and discussions\label{sec:results}}

\subsection{Potential energy surface}

In Figs.~\ref{fig:pes_sm}, \ref{fig:pes_gd}, and \ref{fig:pes_dy}
plotted are the PESs within the
$(\beta,\gamma)$-deformation space for the studied
nuclei $^{148-154}$Sm, $^{150-156}$Gd, and $^{152-158}$Dy,
respectively. 
In these figures, the \mbos, \pbos, and CQF PESs are compared with each other. 
Note that the PESs for the $N=84$ and 96 nuclei in each isotopic chain
have not been plotted in the figures, since they turned out to be
strikingly similar to those for their neighbouring isotopes
with $N=86$ and 94, respectively. 
Here we mainly discuss the PESs for the Sm isotopes, whereas we
confirmed that the main conclusions were basically the same for the Gd
and Dy isotopes.

There is an anzats that the deformation parameter $\beta$ in the IBM can be
related to the one in the geometrical collective model, denoted as
$\bar\beta$, in such a way that they are proportional to each other,
i.e., $\beta=C_\beta\bar\beta$ \cite{ginocchio1980}, where $C_\beta$ 
is the scaling factor and typically takes values $C_\beta\approx 3-5$ in the
rare-earth region \cite{nomura2008}. 
In the \mbos framework, the coefficient $C_\beta$ has been explicitly
determined by the mapping. 
In Figs.~\ref{fig:pes_sm}, \ref{fig:pes_gd}, and \ref{fig:pes_dy},
however, the \mbos PESs are drawn in terms of the $\beta$ 
deformation in the IBM, in order that one can directly compare them
with the \pbos and CQF PESs.

In general, from Figs.~\ref{fig:pes_sm}, \ref{fig:pes_gd}, and
\ref{fig:pes_dy}, the PESs in the \mbos turned out to be more 
strongly deformed and suggested less striking change in topology as
functions of $N$ than those obtained from the \pbos and CQF
Hamiltonians. 
In Fig.~\ref{fig:pes_sm} the \mbos PES for the nucleus
$^{148}$Sm exhibits a nearly spherical mean-field minimum around
$\beta=0.5$. 
In the same figure, one sees that the
location of the minimum, denoted as $\beta_{\rm min}$, in the \mbos PES
jumps from $^{148}$Sm ($\beta_{\rm min}\approx 0.5$) to $^{150}$Sm ($\beta_{\rm min}\approx 1.0$). 
The latter nucleus is suggested to be already well deformed in the \mbos
calculation. For the $^{152,154}$Sm nuclei, one sees even more 
pronounced prolate minimum at $\beta\approx 1.0$, i.e., deeper in energy
in both $\beta$ and $\gamma$ directions, in the corresponding
\mbos PESs. 
On the other hand, the \pbos PESs, depicted in the middle row of
Fig.~\ref{fig:pes_sm}, exhibit a more dramatic change in its 
topology as a function of $N$: spherical
minimum at $\beta=0$ at $^{148}$Sm, weakly prolate deformed minimum at
$^{150}$Sm, softer minimum in both $\beta$ and
$\gamma$ directions at $^{152}$Sm characteristic of the critical-point
nucleus, and well developed prolate minimum at $^{154}$Sm. 
There is no noticeable difference between the PESs obtained from the
\pbos and CQF Hamiltonians.

%
%

\begin{figure}[htb!]
\begin{center}
\includegraphics[width=\linewidth]{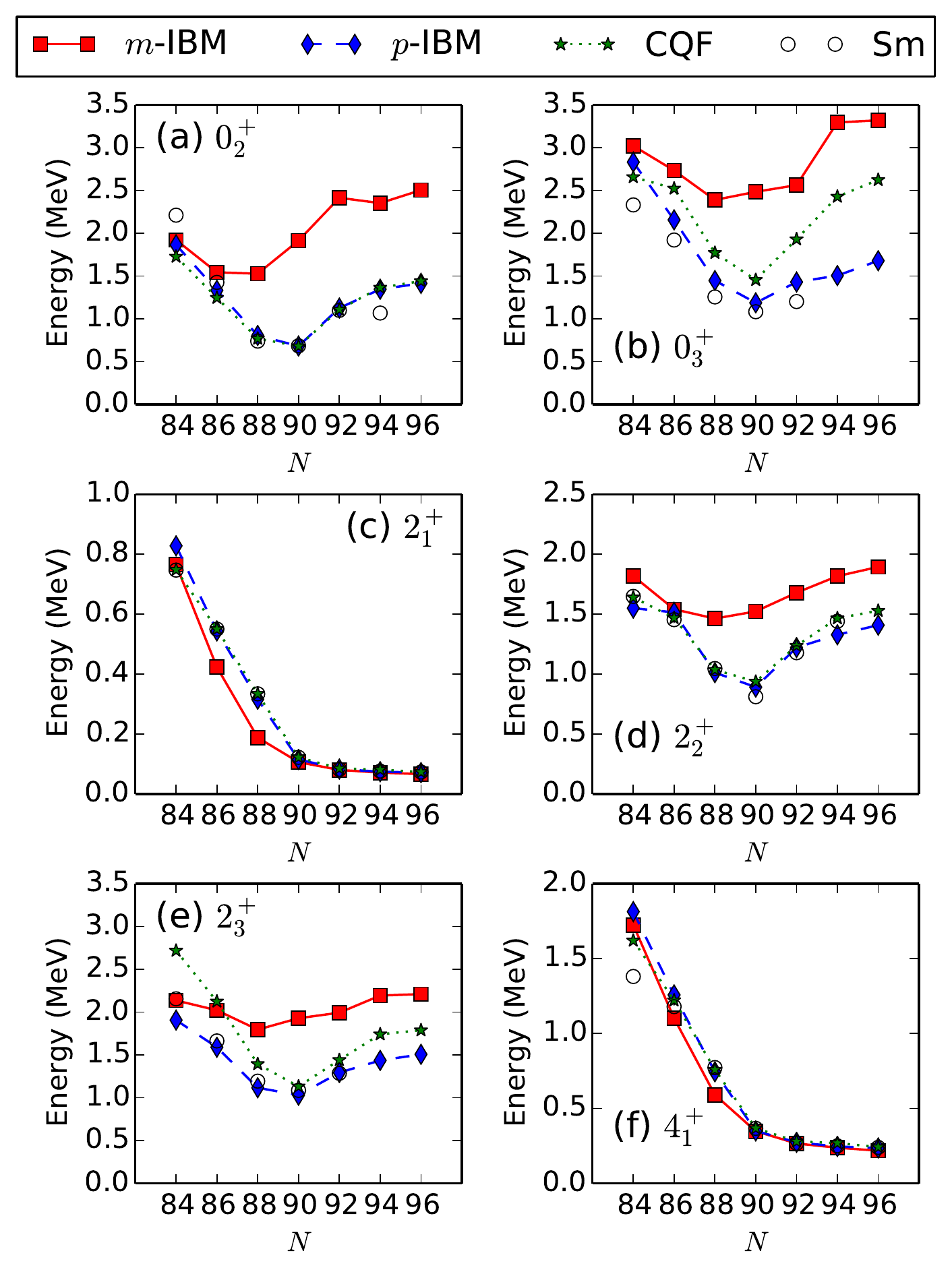} 
\caption{(Color online) Excitation
 energies of the low-lying states $0^+_2$ (a), $0^+_3$ (b), $2^+_1$ (c), $2^+_2$ (d),
 $2^+_3$ (e), and $4^+_1$ (f) for the $^{150-158}$Sm
 isotopes are plotted against the neutron number $N$. The results of the
 three versions of the IBM calculations,  i.e., \mbos, \pbos, and CQF,
 are compared with each other and with the 
 experimental data \cite{data}.}
\label{fig:energies_sm} 
\end{center}
\end{figure}

\begin{figure}[htb!]
\begin{center}
\includegraphics[width=\linewidth]{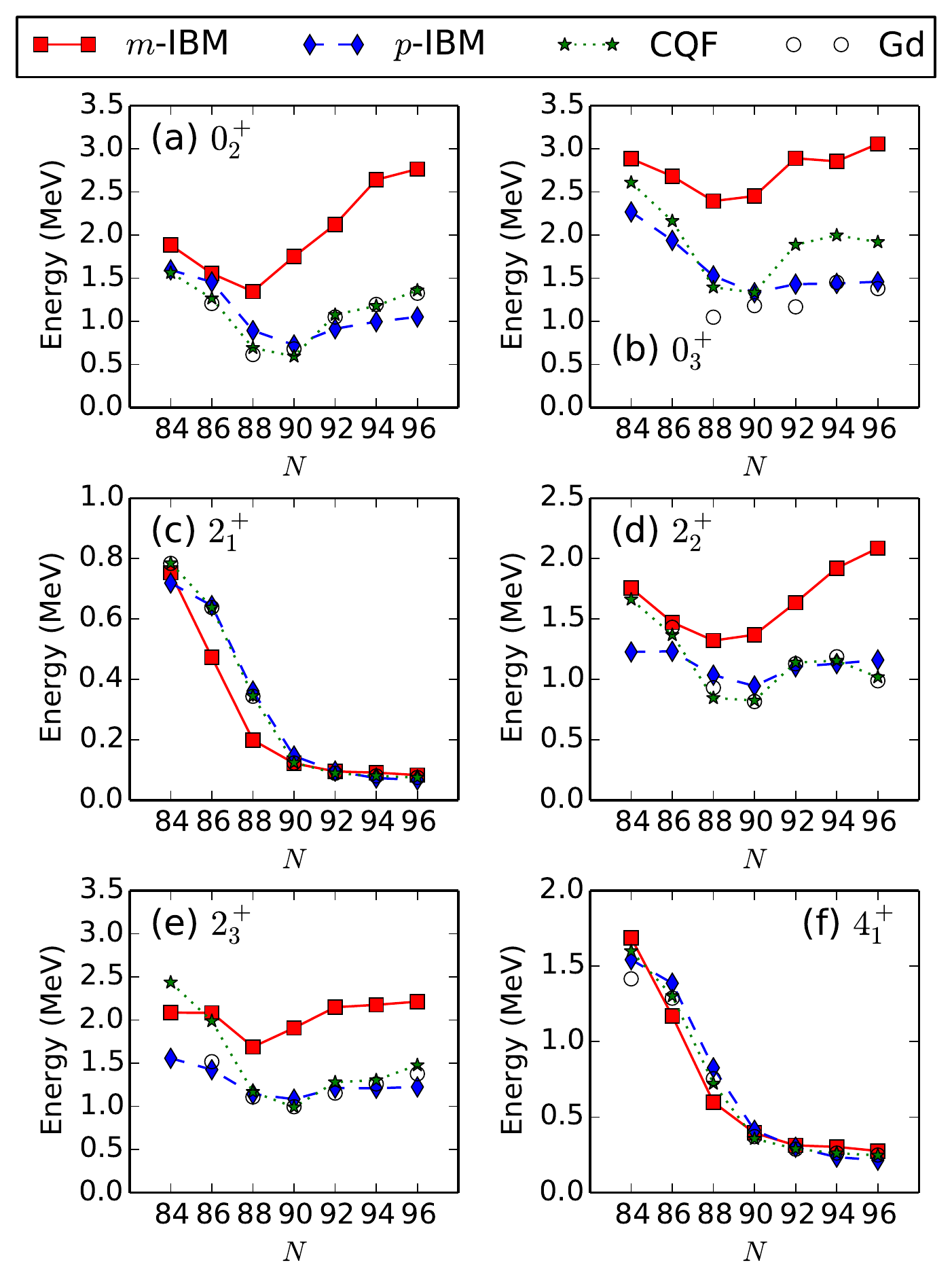} 
\caption{(Color online) Same as Fig.~\ref{fig:energies_sm}, but for the $^{148-160}$Gd isotopes.}
\label{fig:energies_gd} 
\end{center}
\end{figure}

\begin{figure}[htb!]
\begin{center}
\includegraphics[width=\linewidth]{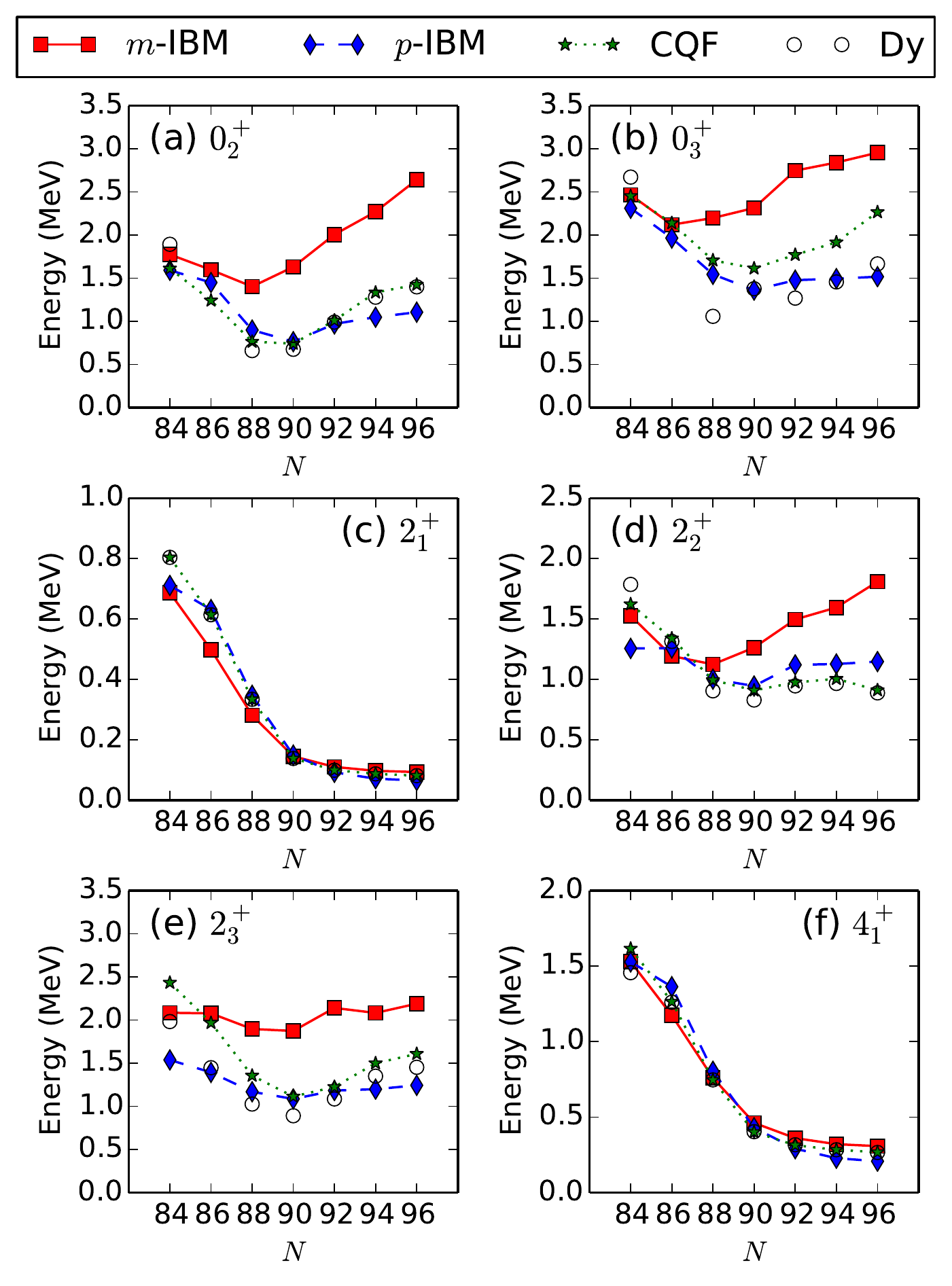} 
\caption{(Color online) Same as Fig.~\ref{fig:energies_sm}, but for the $^{150-162}$Dy isotopes.}
\label{fig:energies_dy} 
\end{center}
\end{figure}

\subsection{Excitation energies}

As a reminder of the results in
Refs.~\cite{nomura2010,kotila2012,zhang2017}, we plotted in Figs.~\ref{fig:energies_sm},
\ref{fig:energies_gd}, and \ref{fig:energies_dy} the excitation energies
of the low-lying states in the $^{146-158}$Sm, $^{148-160}$Gd, and
$^{150-162}$Dy isotopes, respectively, that are 
relevant to the $\tp$ and $\pt$ transfer reactions studied in this work.

\subsubsection{Sm isotopes}

In Fig.~\ref{fig:energies_sm} we display the calculated excitation energies of Sm isotopes. 
The shape phase transition can be identified by the sharp parabolic 
systematics of both the $0^+_2$ and $0^+_3$ energy levels centered
around $N=90$, corresponding to the X(5) critical-point nucleus $^{152}$Sm
\cite{casten2001}.  
But in the \mbos results, the $0^+_2$ and $0^+_3$ energy levels
become lowest rather at $N=88$. 
Both phenomenological (\pbos and CQF) calculations 
reproduced the experimental $0^+_2$ and $0^+_3$ energy levels very
well, while the IBM-2 description looks slightly better than the IBM-1
one. 
In the \mbos, evolution of the energy
levels generally looks more moderate than in the other two calculations. 
Moreover,  both the non-yrast $0^+_{2}$ and $0^+_3$ energies were
overestimated by the \mbos calculation. 
This most likely traces back to the fact that the underlying SCMF PESs suggested a
too deformed mean-field minimum \cite{nomura2010} and that the
corresponding mapped IBM-2 produced a rather rotational energy spectrum. 
Almost the same conclusion as for the results of the non-yrast $0^+$
states can be reached in the comparisons of the $2^+_{2}$
(Fig.~\ref{fig:energies_sm}(d)) and $2^+_3$ (Figs.~\ref{fig:energies_sm}(e))
energy levels.

In Fig.~\ref{fig:energies_sm}(c), the $2^+_1$ energy level has been
reproduced very well by the three calculations. But for the transitional
nuclei, i.e., $^{150}$Sm ($N=88$) and $^{152}$Sm ($N=90$), it has been
predicted to be too low in energy in the \mbos, suggesting rather deformed energy spectra.

As seen from Fig.~\ref{fig:energies_sm}(f), 
the three IBM calculations reproduced very nicely the experimental
$4^+_1$ energy level. However, for the nucleus $^{146}$Sm in particular,
the calculations could not account for the low-lying $4^+_1$ state,
resulting in the predicted energy ratio $R\equiv E(4^+_1)/E(2^+_1)$ that
is below the vibrational limit, i.e., $R<2$. 
This is mainly because of the limited configuration space used in the
present version of the IBM, that is built only on the collective $s$ and
$d$ bosons.

\subsubsection{Gd isotopes}

In the Gd isotopic chain, the experimental $0^+_2$ energy level, shown in
Fig.~\ref{fig:energies_gd}(a), exhibits parabolic behaviour, being
lowest in energy at $N=88$.  
The \mbos result followed this systematics nicely,
but systematically overestimated the data, due to 
the same reasons as we discussed in the previous section. 
The \pbos and CQF calculations provided an excellent description of the
data but, at variance with the \mbos result and the experiment,
suggested that the $0^+_2$ level was lowest at $N=90$.

Compared to the results for the Sm isotopes, as seen from
Fig.~\ref{fig:energies_gd}(b) the experimental $0^+_3$ energy level in
Gd does not show a significant, but rather irregular, 
$N$ dependence for $88\leq N\leq 96$. 
Specially the \pbos calculation reproduced this trend fairly well. 
However, the agreement with the experimental data in the $0^+_3$
excitation energies appears to be not as good as in the $0^+_2$ ones
(Fig.~\ref{fig:energies_gd}(a)), even 
in the phenomenological \pbos and CQF calculations. 
Let us recall that the low-lying $0^+$ excited states in Gd and Dy
isotopes have often been attributed to additional degrees of freedom,
such as intruder excitations, which are beyond the configuration spaces
considered in the present IBM framework.

Both the $2^+_1$ (Fig.~\ref{fig:energies_gd}(c)) and $4^+_1$
(Fig.~\ref{fig:energies_gd}(f)) excitation energies 
have been nicely described by the three calculations. 
As seen from Figs.~\ref{fig:energies_gd}(d) and
\ref{fig:energies_gd}(e), the phenomenological IBM calculations
reproduced the non-yrast $2^+$ levels, but the \mbos overestimated them.

\subsubsection{Dy isotopes}

The main conclusion from the comparisons between 
the theoretical and experimental excitation spectra for the Dy
isotopes in Fig.~\ref{fig:energies_dy} turned out be basically the same as for the
Gd nuclei, discussed in the previous section. 
Namely: 
The \mbos overestimated the experimental data for the non-yrast states,
and differed in the predicted energy-level systematics from the \pbos
and CQF ones; The experimental $0^+_3$ energy level
exhibits rather irregular systematics against $N$, and this experimental
trend was not accounted for by the present version of 
the IBM comprising only collective $s$ and $d$ bosons.

%
%

\begin{figure*}[htb!]
\begin{center}
\includegraphics[width=0.7\linewidth]{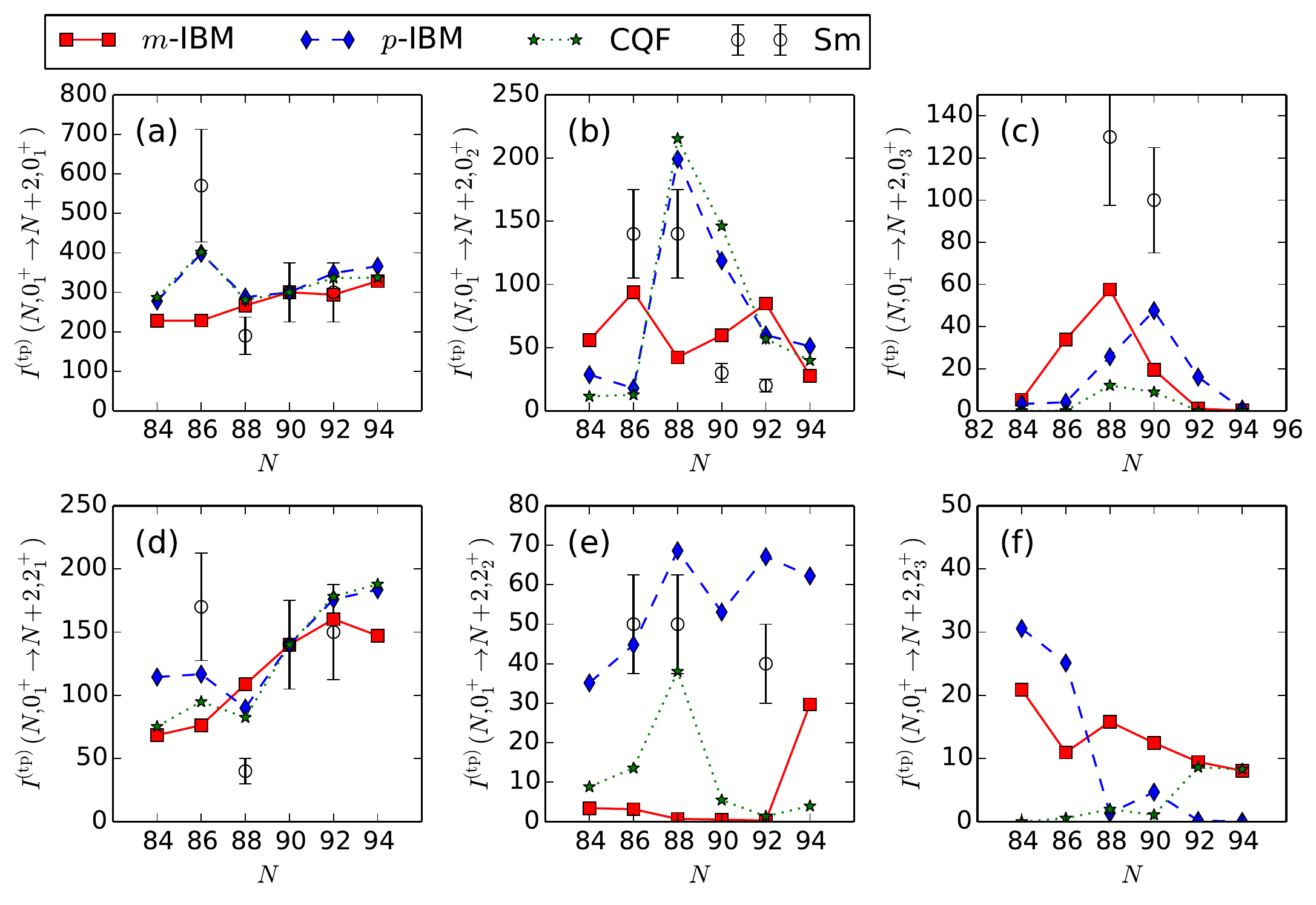} 
\caption{(Color online) The $(t,p)$ transfer reaction intensities for the $^{146-156}$Sm
 isotopes. The results of the $m$-IBM, $p$-IBM, and CQF calculations are
 compared with each other and with the experimental data
 \cite{BJERREGAARD1966145}. 
The scale factors $t_0$ and $t_2$ in the $\tp$ transfer
 operators have been fitted to the experimental data for 
the $0^+_1(^{152}{\rm Sm})\to 0^+_1(^{154}$Sm) and $0^+_1(^{152}{\rm
 Sm})\to 2^+_1(^{154}$Sm) transfer reactions, respectively.} 
\label{fig:tp_sm} 
\end{center}
\end{figure*}

\begin{figure*}[htb!]
\begin{center}
\includegraphics[width=0.7\linewidth]{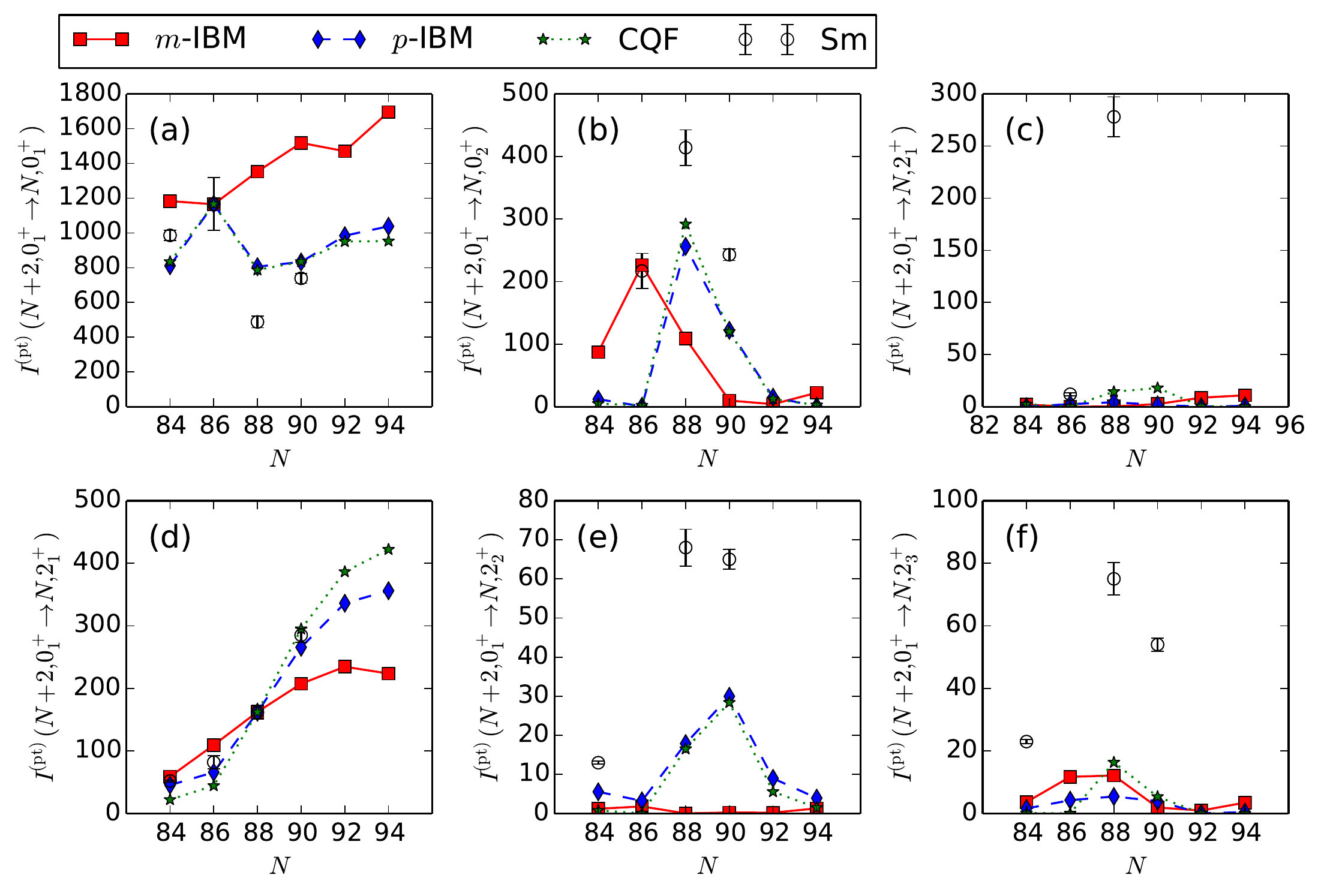} 
\caption{(Color online) Similar to Fig.~\ref{fig:tp_sm}, but for the
 $(p,t)$ transfer reaction
 intensities for the $^{146-156}$Sm
 isotopes. The experimental data have been taken from
 Ref.~\cite{DEBENHAM1972385}. 
The scale factors $t_0$ and $t_2$ in the $\pt$ transfer
 operators have been fitted to the experimental data for 
the $0^+_1(^{150}{\rm Sm})\to 0^+_1(^{148}$Sm) and $0^+_1(^{152}{\rm
 Sm})\to 2^+_1(^{150}$Sm) transfer reactions, respectively.
}
\label{fig:pt_sm} 
\end{center}
\end{figure*}

%
%

\begin{figure*}[htb!]
\begin{center}
\includegraphics[width=0.7\linewidth]{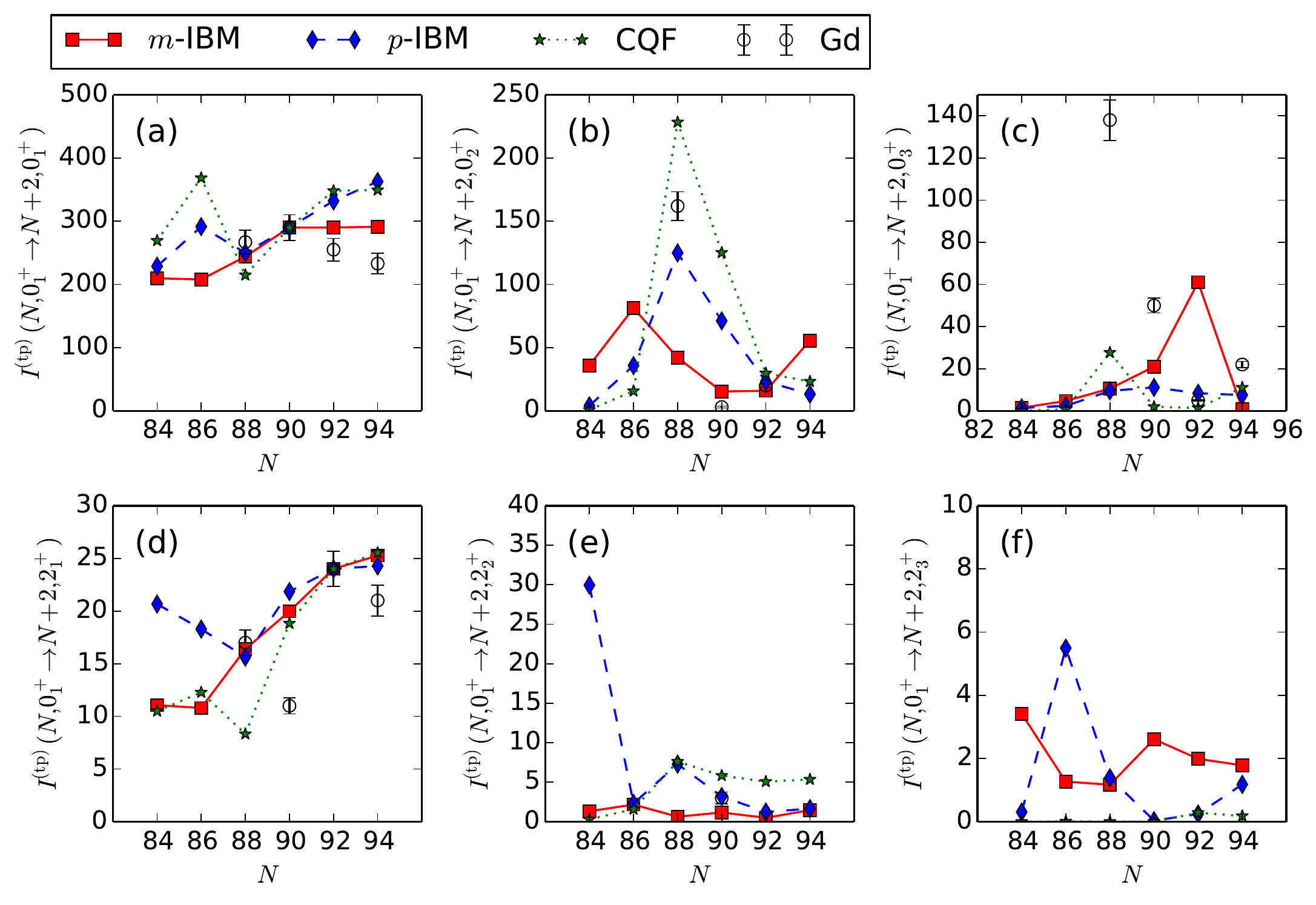} 
\caption{(Color online) Same as Fig.~\ref{fig:tp_sm}, but for the $^{158-158}$Gd
 isotopes. The experimental data have been taken from
 Refs.~\cite{SHAHABUDDIN1980109,LOVHOIDEN1989157,lovhoiden1986}. 
The scale factors $t_0$ and $t_2$ have been fitted to the experimental data for 
the $0^+_1(^{154}{\rm Gd})\to 0^+_1(^{156}$Gd) and $0^+_1(^{156}{\rm
 Gd})\to 2^+_1(^{158}$Gd) transfer reactions, respectively.
}
\label{fig:tp_gd} 
\end{center}
\end{figure*}

\begin{figure*}[htb!]
\begin{center}
\includegraphics[width=0.7\linewidth]{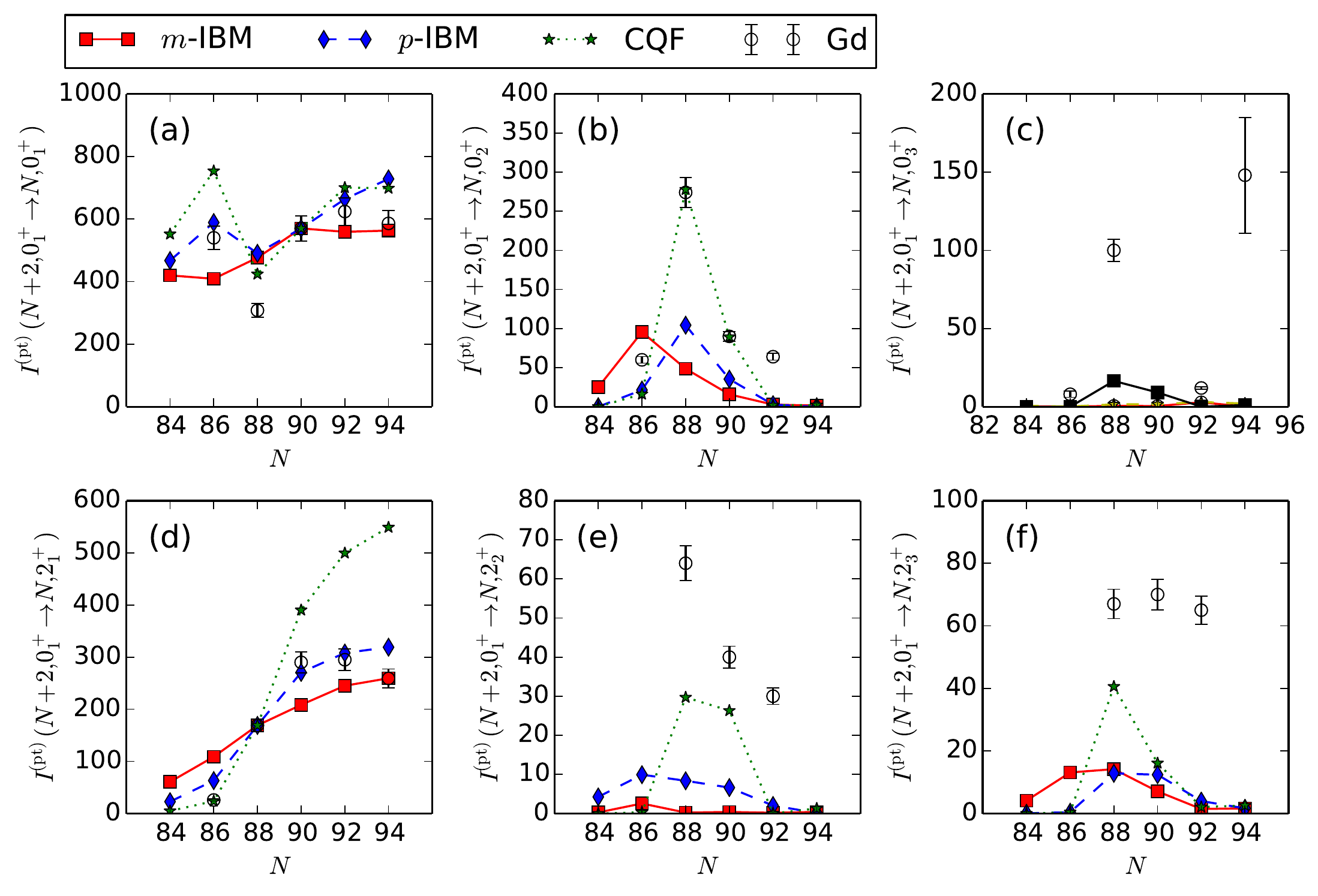} 
\caption{(Color online) Same as Fig.~\ref{fig:pt_sm}, but for the $^{158-158}$Gd
 isotopes. The experimental data have been taken from
 Ref.~\cite{fleming1973}.
The scale factors $t_0$ and $t_2$ have been fitted to the experimental data for 
the $0^+_1(^{156}{\rm Gd})\to 0^+_1(^{154}$Gd) and $0^+_1(^{154}{\rm
 Gd})\to 2^+_1(^{152}$Gd) transfer reactions, respectively.
}
\label{fig:pt_gd} 
\end{center}
\end{figure*}

%
%

\begin{figure*}[htb!]
\begin{center}
\includegraphics[width=0.7\linewidth]{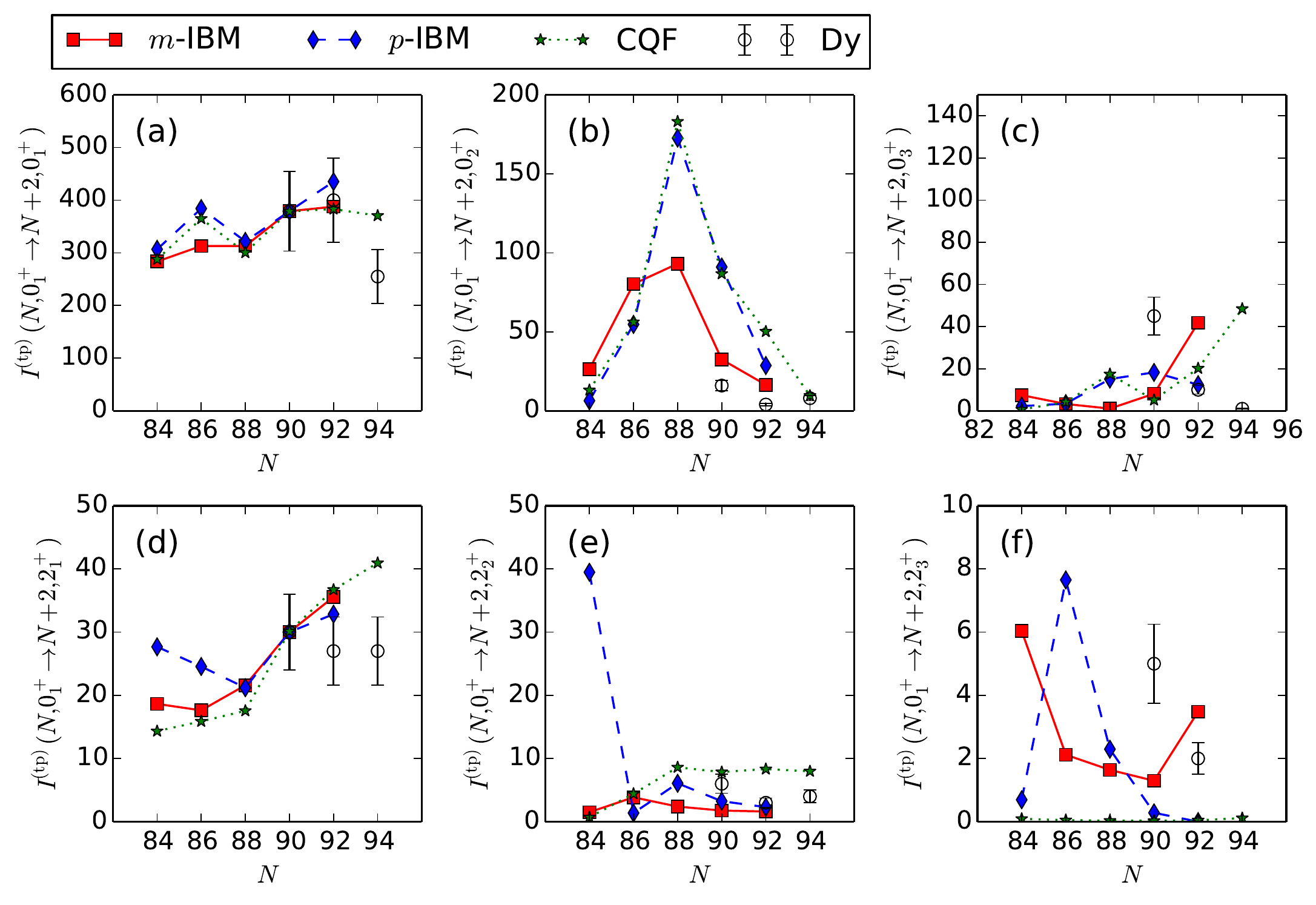} 
\caption{(Color online) Same as Fig.~\ref{fig:tp_sm}, but for the $^{150-160}$Dy
 isotopes. The experimental data have been taken from
 Ref.~\cite{burke1988}. 
The scale factors $t_0$ and $t_2$ have been fitted to the experimental data for 
the $0^+_1(^{156}{\rm Dy})\to 0^+_1(^{158}$Dy) and $0^+_1(^{156}{\rm
 Dy})\to 2^+_1(^{158}$Dy) transfer reactions, respectively.
}
\label{fig:tp_dy} 
\end{center}
\end{figure*}

\begin{figure*}[htb!]
\begin{center}
\includegraphics[width=0.7\linewidth]{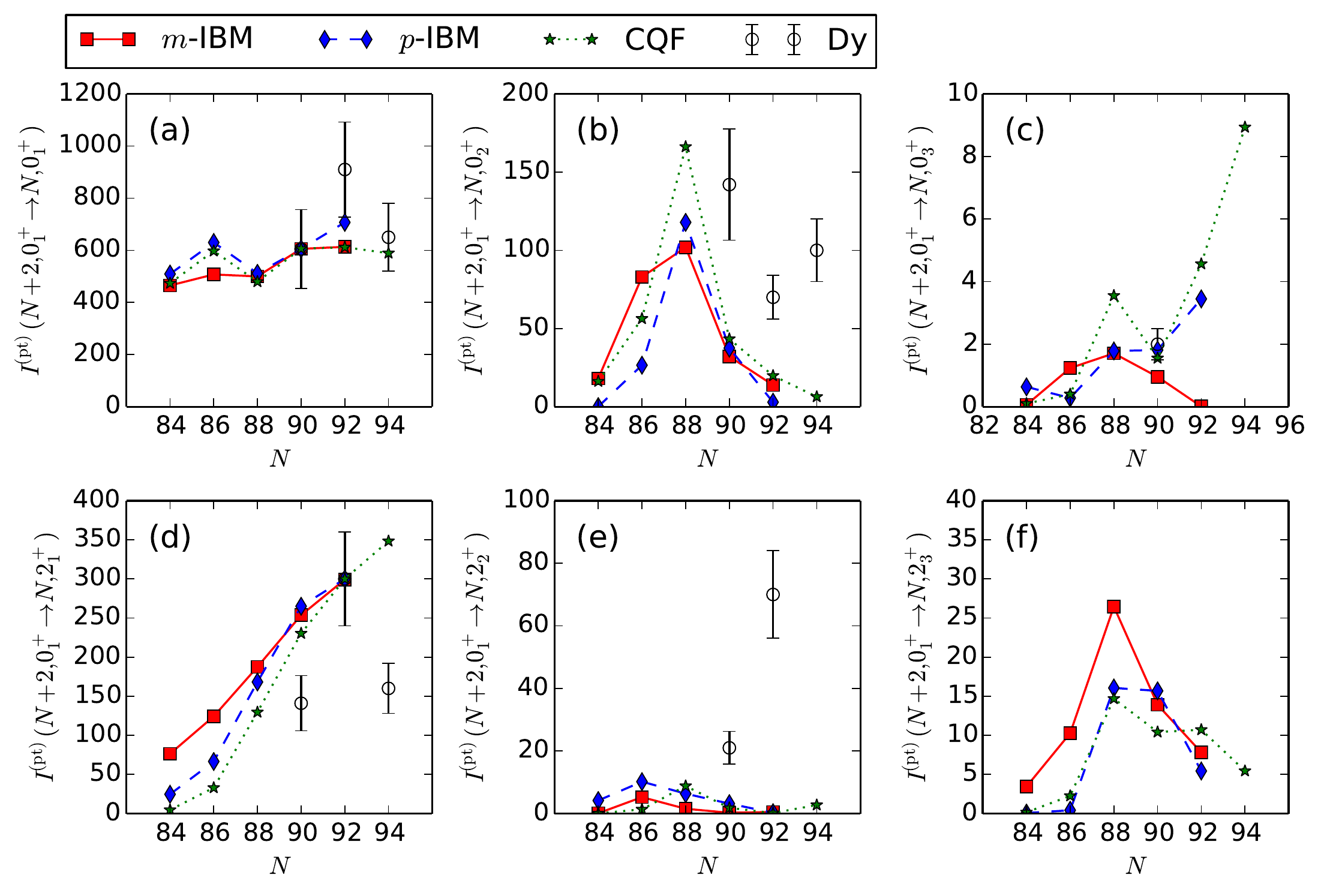} 
\caption{(Color online) Same as Fig.~\ref{fig:pt_sm}, but for the $^{150-160}$Dy
 isotopes. The experimental data have been taken from
 Ref.~\cite{kolata1977} for $^{158}$Dy$\pt^{156}$Dy and
 Ref.~\cite{maher1972} for $^{162}$Dy$\pt^{160}$Dy and
 $^{160}$Dy$\pt^{158}$Dy reactions. 
The scale factors $t_0$ and $t_2$ have been fitted to the experimental data for 
the $0^+_1(^{158}{\rm Dy})\to 0^+_1(^{156}$Dy) and $0^+_1(^{160}{\rm
 Dy})\to 2^+_1(^{158}$Dy) transfer reactions, respectively.
}
\label{fig:pt_dy} 
\end{center}
\end{figure*}

\subsection{$\tp$ and $\pt$ transfer reactions}

Let us now turn to the discussions about the $\tp$ and $\pt$ transfer
reactions. As in the earlier IBM calculations for the two-nucleon transfer
reactions \cite{arima1977trf,scholten1980phd,zhang2017}, we compare the 
calculated $\tp$ and $\pt$ transfer intensities with the 
experimental cross sections measured at particular laboratory angles.
To facilitate the comparisons, we have determined the overall scale factors
$t_0$ and $t_2$ in the transfer operators (see, 
Eqs.~(\ref{eq:opt0}) and (\ref{eq:opt2})) so as to reproduce the experimental 
$0^+_{1,i}\to 0^+_{1,f}$ and $0^+_{1,i}\to 2^+_{1,f}$ transfer reaction
cross sections at given angles for particular nuclei. 
More details are mentioned in the captions to Figs.~\ref{fig:tp_sm} --
\ref{fig:pt_dy}.

\subsubsection{Sm isotopes}

We show in Figs.~\ref{fig:tp_sm} and \ref{fig:pt_sm} the calculated
$\tp$ and $\pt$ transfer reaction intensities for the Sm isotopes as
functions of $N$. 
The experimental data, available in
Refs.~\cite{BJERREGAARD1966145,DEBENHAM1972385}, are also included in
the plot. 

In Fig.~\ref{fig:tp_sm}, for many of the $\tp$ transfer reactions the
$m$-IBM results exhibit a 
certain discontinuity around particular nucleus in the transitional region. 
In general, the $\tp$ reaction rates resulting from the $m$-IBM did not
exhibit change with $N$ as rapid as those from the \pbos and CQF and, in
some reactions, show completely different $N$ dependence from the
latter. A typical example is the $I^{\rm (tp)}(N,0^+_1\to
N+2,0^+_2)$ reaction rate (see, Fig.~\ref{fig:tp_sm}(b)). 
The difference between the microscopic and phenomenological IBM
calculations in the nature of the structural evolution is consistent
with what we observed in the PESs (see, Fig.~\ref{fig:pes_sm}) and
excitation energies (Fig.~\ref{fig:energies_sm}). 
The two phenomenological calculations, i.e., \pbos and CQF, have
provided similar results to each other both qualitatively and
quantitatively. 
We note that both the theoretical $I^{\rm (tp)}(N,0^+_1\to N+2,0^+_1)$ and $I^{\rm
(tp)}(N,0^+_1\to N+2,2^+_1)$ intensities from the \pbos and CQF
calculations are in a very good agreement with the corresponding experimental data.

Also for the $\pt$ transfer reactions in Fig.~\ref{fig:pt_sm}, the \mbos
calculation indicated that the phase transition occurred more moderately
than in the \pbos and CQF results and was, in some reactions, observed
at somewhat different neutron number than the \pbos and CQF results
(see, e.g., Fig.~\ref{fig:pt_sm}(b)).

\subsubsection{Gd isotopes}

In Fig.~\ref{fig:tp_gd} we plotted the theoretical $\tp$ transfer reaction
intensities for the Gd isotopes, in comparison with the experimental
data available at \cite{SHAHABUDDIN1980109,LOVHOIDEN1989157,lovhoiden1986}. 
In most of the considered $\tp$ reactions, the \mbos calculation
indicates an irregular behaviour with $N$, suggesting the rapid shape
transition. 
However, the location at which such an irregularity appears in the \mbos
results is at variance with the \pbos and CQF results in the $\tp$
transfer intensities $I^{\rm (tp)}(N,0^+_1\to N+2,0^+_2)$ (Fig.~\ref{fig:tp_gd}(b)), $I^{\rm (tp)}(N,0^+_1\to
N+2,0^+_3)$ (Fig.~\ref{fig:tp_gd}(c)), and $I^{\rm (tp)}(N,0^+_1\to
N+2,2^+_3)$ (Fig.~\ref{fig:tp_gd}(f)). 
All the three IBM calculations commonly failed to reproduce the
experimental data for the $0^+_1\to 0^+_3$ $\tp$ transfer reactions. 
This confirms that the $0^+_3$ state could be well beyond the model
space of the $sd$-IBM, which corroborates with the comparisons of the
excitation energies for the same state. 
Those correlations that are out of the IBM space could be effectively
taken into account by the inclusion of higher-order terms in the
transfer operators in Eqs.~(\ref{eq:opt0}) and (\ref{eq:opt2}), but such
an extension would involve additional parameters to be determined and is
beyond the scope of the present study.

One sees in Fig.~\ref{fig:tp_gd}(e) an anomalously large difference in
the $I^{\rm (tp)}(N,0^+_1\to N+2,2^+_2)$ values calculated within the \pbos between
$^{148}$Gd ($N=84$) and $^{150}$Gd ($N=86$). 
This could be a consequence of the fact that the present \pbos
calculation, perhaps due to a poor fit to the experimental spectra or
some missing correlations, did not describe well the $2^+_2$ excitation energy at the nucleus
$^{148}$Gd (see, Fig.~\ref{fig:energies_gd}(d)). 
As we show below in Fig.~\ref{fig:tp_dy}(e), the same problem was
observed in the $\tp$ reactions in the Dy nuclei.

As seen from the results for the $\pt$ transfer reaction intensities shown in
Fig.~\ref{fig:pt_gd}, the three IBM calculations consistently point to an abrupt
change around the transitional nucleus $^{152}$Gd ($N=88$) or $^{154}$Gd ($N=90$). 
On the other hand, notable discrepancy is found between the theoretical
$I^{\rm (pt)}(N+2,0^+_1\to N,0^+_3)$ (Fig.~\ref{fig:pt_gd}(c))
and $I^{\rm (pt)}(N+2,0^+_1\to N,2^+_3)$ (Fig.~\ref{fig:pt_gd}(f)) 
intensities and the corresponding experimental data. 
As we have already observed, the \mbos result appears to suggest a more
moderate nuclear structural evolution with $N$ than the \pbos and CQF ones.

\subsubsection{Dy isotopes}

The calculated $\tp$ transfer reaction intensities for the Dy
isotopes are plotted in Fig.~\ref{fig:tp_dy}. 
In the present IBM-2
(both \mbos and \pbos) calculations, however, the heaviest nucleus 
$^{162}$Dy turned out be beyond the limit of the current version of the computer
program, and were not plotted in the figure, as well as in the following
Fig.~\ref{fig:pt_dy}. 
In all the three IBM calculations, a discontinuity of the $\tp$ transfer
intensities has been suggested in the transitional nuclei with $N\approx 90$,
that is a clear signature of the shape phase transition. 
It is remarkable that, compared to the Sm and Gd results
(Figs.~\ref{fig:tp_sm}--\ref{fig:pt_gd}), the three different IBM 
calculations for the Dy isotopes provided results very much similar to each other both
at qualitative and quantitative levels, except
perhaps for the $I^{\rm (tp)}(N,0^+_1\to N+2,2^+_3)$ intensity (Fig.~\ref{fig:tp_dy}(f)). 
The above observation holds, to a greater extent, for the $\pt$ transfer
reactions in Fig.~\ref{fig:pt_dy}.

\section{Summary\label{sec:summary}}

The interacting boson model, that is based on the
microscopic framework of the self-consistent
mean-field method, has been applied to study the two-nucleon transfer
reactions as a signature of the shape phase transition. 
Constrained SCMF calculations have been performed within the
Hartree-Fock plus BCS method based on the Skyrme energy density
functional to provide a microscopic
input to completely determine the Hamiltonian of the IBM-2. 
The $\tp$ and $\pt$ transfer
reaction intensities for the rare-earth nuclei $^{146-158}$Sm,
$^{148-160}$Gd, and $^{150-162}$Dy, which 
are an excellent example of the spherical-to-axially-deformed shape
phase transition, have been computed by using the wave functions of the
mapped IBM-2 Hamiltonian. 
Apart from the overall scaling factors for the transfer operators 
constant for each isotopic chain, no phenomenological adjustment has been
made. 
The $\tp$ and $\pt$ transfer reaction intensities calculated by the 
microscopically-formulated IBM-2 have been compared with
the results from the 
purely phenomenological IBM-2 and IBM-1 with parameters determined
by the fits to experimental excitation spectra in each nucleus.

The overall systematic behaviors of the calculated $\tp$ and $\pt$ transfer
reaction intensities against the neutron number showed that
the shape transition occurred more moderately in the microscopic 
IBM than was suggested by the phenomenological IBM. 
This finding corroborates with the quantitative, as well as the
qualitative, differences in the predictions of the low-lying energy
levels between the microscopic and 
phenomenological calculations. 
Such differences seem to have originated from the
SCMF calculation of the PESs with a specific choice
of the energy density functional, which suggested that
the nuclear structure evolution took place more moderately than was
expected in phenomenological models.

However, all the three IBM calculations consistently pointed to an irregular
behaviour of the $\tp$ and $\pt$ transfer reaction intensities at
specific neutron numbers, and indicated that the
two-neutron transfer reactions can be used as a signature of the shape
phase transitions. The results presented in this paper also confirmed
that the SCMF-to-IBM mapping procedure was a sound approach to the simultaneous
description of the decay spectroscopy in a single nucleus and the transfer reactions
between different nuclei.

\acknowledgments
This work was supported in part by the QuantiXLie Centre of Excellence, a project
co-financed by the Croatian Government and European Union through the
European Regional Development Fund - the Competitiveness and Cohesion
Operational Programme (Grant KK.01.1.1.01.0004).
One of us (Y.Z.) acknowledges support from the Natural Science
Foundation of China (Grant No. 11875158).

\bibliography{refs}

\begin{thebibliography}{40}
\expandafter\ifx\csname natexlab\endcsname\relax\def\natexlab#1{#1}\fi
\expandafter\ifx\csname bibnamefont\endcsname\relax
  \def\bibnamefont#1{#1}\fi
\expandafter\ifx\csname bibfnamefont\endcsname\relax
  \def\bibfnamefont#1{#1}\fi
\expandafter\ifx\csname citenamefont\endcsname\relax
  \def\citenamefont#1{#1}\fi
\expandafter\ifx\csname url\endcsname\relax
  \def\url#1{\texttt{#1}}\fi
\expandafter\ifx\csname urlprefix\endcsname\relax\def\urlprefix{URL }\fi
\providecommand{\bibinfo}[2]{#2}
\providecommand{\eprint}[2][]{\url{#2}}

\bibitem[{\citenamefont{Cejnar and Jolie}(2009)}]{cejnar2009}
\bibinfo{author}{\bibfnamefont{P.}~\bibnamefont{Cejnar}} \bibnamefont{and}
  \bibinfo{author}{\bibfnamefont{J.}~\bibnamefont{Jolie}},
  \bibinfo{journal}{Progress in Particle and Nuclear Physics}
  \textbf{\bibinfo{volume}{62}}, \bibinfo{pages}{210 } (\bibinfo{year}{2009}),
  ISSN \bibinfo{issn}{0146-6410},
  \urlprefix\url{http://www.sciencedirect.com/science/article/pii/S0146641008000719}.

\bibitem[{\citenamefont{Cejnar et~al.}(2010)\citenamefont{Cejnar, Jolie, and
  Casten}}]{cejnar2010}
\bibinfo{author}{\bibfnamefont{P.}~\bibnamefont{Cejnar}},
  \bibinfo{author}{\bibfnamefont{J.}~\bibnamefont{Jolie}}, \bibnamefont{and}
  \bibinfo{author}{\bibfnamefont{R.~F.} \bibnamefont{Casten}},
  \bibinfo{journal}{Rev. Mod. Phys.} \textbf{\bibinfo{volume}{82}},
  \bibinfo{pages}{2155} (\bibinfo{year}{2010}).

\bibitem[{\citenamefont{Iachello}(2011)}]{iachello2011a}
\bibinfo{author}{\bibfnamefont{F.}~\bibnamefont{Iachello}},
  \bibinfo{journal}{Revista Nuovo Cimento} \textbf{\bibinfo{volume}{34}},
  \bibinfo{pages}{617} (\bibinfo{year}{2011}).

\bibitem[{\citenamefont{Carr}(2010)}]{carr-book}
\bibinfo{editor}{\bibfnamefont{L.}~\bibnamefont{Carr}}, ed.,
  \emph{\bibinfo{title}{Understanding Quantum Phase Transitions}}
  (\bibinfo{publisher}{CRC Press}, \bibinfo{year}{2010}).

\bibitem[{\citenamefont{Maxwell et~al.}(1966)\citenamefont{Maxwell, Reynolds,
  and Hintz}}]{maxwell1966}
\bibinfo{author}{\bibfnamefont{J.~R.} \bibnamefont{Maxwell}},
  \bibinfo{author}{\bibfnamefont{G.~M.} \bibnamefont{Reynolds}},
  \bibnamefont{and} \bibinfo{author}{\bibfnamefont{N.~M.} \bibnamefont{Hintz}},
  \bibinfo{journal}{Phys. Rev.} \textbf{\bibinfo{volume}{151}},
  \bibinfo{pages}{1000} (\bibinfo{year}{1966}),
  \urlprefix\url{https://link.aps.org/doi/10.1103/PhysRev.151.1000}.

\bibitem[{\citenamefont{Bjerregaard et~al.}(1966)\citenamefont{Bjerregaard,
  Hansen, Nathan, and Hinds}}]{BJERREGAARD1966145}
\bibinfo{author}{\bibfnamefont{J.}~\bibnamefont{Bjerregaard}},
  \bibinfo{author}{\bibfnamefont{O.}~\bibnamefont{Hansen}},
  \bibinfo{author}{\bibfnamefont{O.}~\bibnamefont{Nathan}}, \bibnamefont{and}
  \bibinfo{author}{\bibfnamefont{S.}~\bibnamefont{Hinds}},
  \bibinfo{journal}{Nuclear Physics} \textbf{\bibinfo{volume}{86}},
  \bibinfo{pages}{145 } (\bibinfo{year}{1966}), ISSN \bibinfo{issn}{0029-5582},
  \urlprefix\url{http://www.sciencedirect.com/science/article/pii/0029558266902975}.

\bibitem[{\citenamefont{Fleming et~al.}(1971)\citenamefont{Fleming, G\"unther,
  Hagemann, Herskind, and Tj\o{}m}}]{fleming1971}
\bibinfo{author}{\bibfnamefont{D.~G.} \bibnamefont{Fleming}},
  \bibinfo{author}{\bibfnamefont{C.}~\bibnamefont{G\"unther}},
  \bibinfo{author}{\bibfnamefont{G.~B.} \bibnamefont{Hagemann}},
  \bibinfo{author}{\bibfnamefont{B.}~\bibnamefont{Herskind}}, \bibnamefont{and}
  \bibinfo{author}{\bibfnamefont{P.~O.} \bibnamefont{Tj\o{}m}},
  \bibinfo{journal}{Phys. Rev. Lett.} \textbf{\bibinfo{volume}{27}},
  \bibinfo{pages}{1235} (\bibinfo{year}{1971}),
  \urlprefix\url{https://link.aps.org/doi/10.1103/PhysRevLett.27.1235}.

\bibitem[{\citenamefont{Casten et~al.}(1972)\citenamefont{Casten, Flynn,
  Hansen, and Mulligan}}]{casten1972}
\bibinfo{author}{\bibfnamefont{R.}~\bibnamefont{Casten}},
  \bibinfo{author}{\bibfnamefont{E.}~\bibnamefont{Flynn}},
  \bibinfo{author}{\bibfnamefont{O.}~\bibnamefont{Hansen}}, \bibnamefont{and}
  \bibinfo{author}{\bibfnamefont{T.}~\bibnamefont{Mulligan}},
  \bibinfo{journal}{Nuclear Physics A} \textbf{\bibinfo{volume}{184}},
  \bibinfo{pages}{357 } (\bibinfo{year}{1972}), ISSN \bibinfo{issn}{0375-9474},
  \urlprefix\url{http://www.sciencedirect.com/science/article/pii/0375947472904149}.

\bibitem[{\citenamefont{Fleming et~al.}(1973)\citenamefont{Fleming, G\"unther,
  Hagemann, Herskind, and Tj\o{}m}}]{fleming1973}
\bibinfo{author}{\bibfnamefont{D.~G.} \bibnamefont{Fleming}},
  \bibinfo{author}{\bibfnamefont{C.}~\bibnamefont{G\"unther}},
  \bibinfo{author}{\bibfnamefont{G.}~\bibnamefont{Hagemann}},
  \bibinfo{author}{\bibfnamefont{B.}~\bibnamefont{Herskind}}, \bibnamefont{and}
  \bibinfo{author}{\bibfnamefont{P.~O.} \bibnamefont{Tj\o{}m}},
  \bibinfo{journal}{Phys. Rev. C} \textbf{\bibinfo{volume}{8}},
  \bibinfo{pages}{806} (\bibinfo{year}{1973}),
  \urlprefix\url{https://link.aps.org/doi/10.1103/PhysRevC.8.806}.

\bibitem[{\citenamefont{Shahabuddin et~al.}(1980)\citenamefont{Shahabuddin,
  Burke, Nowikow, and Waddington}}]{SHAHABUDDIN1980109}
\bibinfo{author}{\bibfnamefont{M.}~\bibnamefont{Shahabuddin}},
  \bibinfo{author}{\bibfnamefont{D.}~\bibnamefont{Burke}},
  \bibinfo{author}{\bibfnamefont{I.}~\bibnamefont{Nowikow}}, \bibnamefont{and}
  \bibinfo{author}{\bibfnamefont{J.}~\bibnamefont{Waddington}},
  \bibinfo{journal}{Nuclear Physics A} \textbf{\bibinfo{volume}{340}},
  \bibinfo{pages}{109 } (\bibinfo{year}{1980}), ISSN \bibinfo{issn}{0375-9474},
  \urlprefix\url{http://www.sciencedirect.com/science/article/pii/0375947480903255}.

\bibitem[{\citenamefont{L\o{}vh\o{}iden
  et~al.}(1986)\citenamefont{L\o{}vh\o{}iden, Thorsteinsen, and
  Burke}}]{lovhoiden1986}
\bibinfo{author}{\bibfnamefont{G.}~\bibnamefont{L\o{}vh\o{}iden}},
  \bibinfo{author}{\bibfnamefont{T.~F.} \bibnamefont{Thorsteinsen}},
  \bibnamefont{and} \bibinfo{author}{\bibfnamefont{D.~G.} \bibnamefont{Burke}},
  \bibinfo{journal}{Physica Scripta} \textbf{\bibinfo{volume}{34}},
  \bibinfo{pages}{691} (\bibinfo{year}{1986}),
  \urlprefix\url{http://stacks.iop.org/1402-4896/34/i=6A/a=025}.

\bibitem[{\citenamefont{L\o{}vh\o{}iden
  et~al.}(1989)\citenamefont{L\o{}vh\o{}iden, Thorsteinsen, Andersen, Kiziltan,
  and Burke}}]{LOVHOIDEN1989157}
\bibinfo{author}{\bibfnamefont{G.}~\bibnamefont{L\o{}vh\o{}iden}},
  \bibinfo{author}{\bibfnamefont{T.}~\bibnamefont{Thorsteinsen}},
  \bibinfo{author}{\bibfnamefont{E.}~\bibnamefont{Andersen}},
  \bibinfo{author}{\bibfnamefont{M.}~\bibnamefont{Kiziltan}}, \bibnamefont{and}
  \bibinfo{author}{\bibfnamefont{D.}~\bibnamefont{Burke}},
  \bibinfo{journal}{Nuclear Physics A} \textbf{\bibinfo{volume}{494}},
  \bibinfo{pages}{157 } (\bibinfo{year}{1989}), ISSN \bibinfo{issn}{0375-9474},
  \urlprefix\url{http://www.sciencedirect.com/science/article/pii/0375947489900171}.

\bibitem[{\citenamefont{Lesher et~al.}(2002)\citenamefont{Lesher, Aprahamian,
  Trache, Oros-Peusquens, Deyliz, Gollwitzer, Hertenberger, Valnion, and
  Graw}}]{lesher2002}
\bibinfo{author}{\bibfnamefont{S.~R.} \bibnamefont{Lesher}},
  \bibinfo{author}{\bibfnamefont{A.}~\bibnamefont{Aprahamian}},
  \bibinfo{author}{\bibfnamefont{L.}~\bibnamefont{Trache}},
  \bibinfo{author}{\bibfnamefont{A.}~\bibnamefont{Oros-Peusquens}},
  \bibinfo{author}{\bibfnamefont{S.}~\bibnamefont{Deyliz}},
  \bibinfo{author}{\bibfnamefont{A.}~\bibnamefont{Gollwitzer}},
  \bibinfo{author}{\bibfnamefont{R.}~\bibnamefont{Hertenberger}},
  \bibinfo{author}{\bibfnamefont{B.~D.} \bibnamefont{Valnion}},
  \bibnamefont{and} \bibinfo{author}{\bibfnamefont{G.}~\bibnamefont{Graw}},
  \bibinfo{journal}{Phys. Rev. C} \textbf{\bibinfo{volume}{66}},
  \bibinfo{pages}{051305} (\bibinfo{year}{2002}),
  \urlprefix\url{https://link.aps.org/doi/10.1103/PhysRevC.66.051305}.

\bibitem[{\citenamefont{Fossion et~al.}(2007)\citenamefont{Fossion, Alonso,
  Arias, Fortunato, and Vitturi}}]{fossion2007}
\bibinfo{author}{\bibfnamefont{R.}~\bibnamefont{Fossion}},
  \bibinfo{author}{\bibfnamefont{C.~E.} \bibnamefont{Alonso}},
  \bibinfo{author}{\bibfnamefont{J.~M.} \bibnamefont{Arias}},
  \bibinfo{author}{\bibfnamefont{L.}~\bibnamefont{Fortunato}},
  \bibnamefont{and} \bibinfo{author}{\bibfnamefont{A.}~\bibnamefont{Vitturi}},
  \bibinfo{journal}{Phys. Rev. C} \textbf{\bibinfo{volume}{76}},
  \bibinfo{pages}{014316} (\bibinfo{year}{2007}),
  \urlprefix\url{https://link.aps.org/doi/10.1103/PhysRevC.76.014316}.

\bibitem[{\citenamefont{Clark et~al.}(2009)\citenamefont{Clark, Casten,
  Bettermann, and Winkler}}]{clark2009}
\bibinfo{author}{\bibfnamefont{R.~M.} \bibnamefont{Clark}},
  \bibinfo{author}{\bibfnamefont{R.~F.} \bibnamefont{Casten}},
  \bibinfo{author}{\bibfnamefont{L.}~\bibnamefont{Bettermann}},
  \bibnamefont{and} \bibinfo{author}{\bibfnamefont{R.}~\bibnamefont{Winkler}},
  \bibinfo{journal}{Phys. Rev. C} \textbf{\bibinfo{volume}{80}},
  \bibinfo{pages}{011303} (\bibinfo{year}{2009}),
  \urlprefix\url{https://link.aps.org/doi/10.1103/PhysRevC.80.011303}.

\bibitem[{\citenamefont{Zhang and Iachello}(2017)}]{zhang2017}
\bibinfo{author}{\bibfnamefont{Y.}~\bibnamefont{Zhang}} \bibnamefont{and}
  \bibinfo{author}{\bibfnamefont{F.}~\bibnamefont{Iachello}},
  \bibinfo{journal}{Phys. Rev. C} \textbf{\bibinfo{volume}{95}},
  \bibinfo{pages}{034306} (\bibinfo{year}{2017}),
  \urlprefix\url{https://link.aps.org/doi/10.1103/PhysRevC.95.034306}.

\bibitem[{\citenamefont{Iachello and Arima}(1987)}]{IBM}
\bibinfo{author}{\bibfnamefont{F.}~\bibnamefont{Iachello}} \bibnamefont{and}
  \bibinfo{author}{\bibfnamefont{A.}~\bibnamefont{Arima}},
  \emph{\bibinfo{title}{The interacting boson model}}
  (\bibinfo{publisher}{Cambridge University Press, Cambridge},
  \bibinfo{year}{1987}).

\bibitem[{\citenamefont{Otsuka et~al.}(1978{\natexlab{a}})\citenamefont{Otsuka,
  Arima, Iachello, and Talmi}}]{OAIT}
\bibinfo{author}{\bibfnamefont{T.}~\bibnamefont{Otsuka}},
  \bibinfo{author}{\bibfnamefont{A.}~\bibnamefont{Arima}},
  \bibinfo{author}{\bibfnamefont{F.}~\bibnamefont{Iachello}}, \bibnamefont{and}
  \bibinfo{author}{\bibfnamefont{I.}~\bibnamefont{Talmi}},
  \bibinfo{journal}{Phys. Lett. B} \textbf{\bibinfo{volume}{76}},
  \bibinfo{pages}{139 } (\bibinfo{year}{1978}{\natexlab{a}}).

\bibitem[{\citenamefont{Otsuka et~al.}(1978{\natexlab{b}})\citenamefont{Otsuka,
  Arima, and Iachello}}]{OAI}
\bibinfo{author}{\bibfnamefont{T.}~\bibnamefont{Otsuka}},
  \bibinfo{author}{\bibfnamefont{A.}~\bibnamefont{Arima}}, \bibnamefont{and}
  \bibinfo{author}{\bibfnamefont{F.}~\bibnamefont{Iachello}},
  \bibinfo{journal}{Nucl. Phys. A} \textbf{\bibinfo{volume}{309}},
  \bibinfo{pages}{1} (\bibinfo{year}{1978}{\natexlab{b}}).

\bibitem[{\citenamefont{Mizusaki and Otsuka}(1996)}]{mizusaki1997}
\bibinfo{author}{\bibfnamefont{T.}~\bibnamefont{Mizusaki}} \bibnamefont{and}
  \bibinfo{author}{\bibfnamefont{T.}~\bibnamefont{Otsuka}},
  \bibinfo{journal}{Prog. Theor. Phys. Suppl.} \textbf{\bibinfo{volume}{125}},
  \bibinfo{pages}{97} (\bibinfo{year}{1996}).

\bibitem[{\citenamefont{Nomura et~al.}(2008)\citenamefont{Nomura, Shimizu, and
  Otsuka}}]{nomura2008}
\bibinfo{author}{\bibfnamefont{K.}~\bibnamefont{Nomura}},
  \bibinfo{author}{\bibfnamefont{N.}~\bibnamefont{Shimizu}}, \bibnamefont{and}
  \bibinfo{author}{\bibfnamefont{T.}~\bibnamefont{Otsuka}},
  \bibinfo{journal}{Phys. Rev. Lett.} \textbf{\bibinfo{volume}{101}},
  \bibinfo{pages}{142501} (\bibinfo{year}{2008}).

\bibitem[{\citenamefont{Nomura et~al.}(2011)\citenamefont{Nomura, Otsuka,
  Shimizu, and Guo}}]{nomura2011rot}
\bibinfo{author}{\bibfnamefont{K.}~\bibnamefont{Nomura}},
  \bibinfo{author}{\bibfnamefont{T.}~\bibnamefont{Otsuka}},
  \bibinfo{author}{\bibfnamefont{N.}~\bibnamefont{Shimizu}}, \bibnamefont{and}
  \bibinfo{author}{\bibfnamefont{L.}~\bibnamefont{Guo}},
  \bibinfo{journal}{Phys. Rev. C} \textbf{\bibinfo{volume}{83}},
  \bibinfo{pages}{041302} (\bibinfo{year}{2011}).

\bibitem[{\citenamefont{Ginocchio and Kirson}(1980)}]{ginocchio1980}
\bibinfo{author}{\bibfnamefont{J.~N.} \bibnamefont{Ginocchio}}
  \bibnamefont{and} \bibinfo{author}{\bibfnamefont{M.~W.}
  \bibnamefont{Kirson}}, \bibinfo{journal}{Nucl. Phys. A}
  \textbf{\bibinfo{volume}{350}}, \bibinfo{pages}{31} (\bibinfo{year}{1980}).

\bibitem[{\citenamefont{Arima and Iachello}(1977)}]{arima1977trf}
\bibinfo{author}{\bibfnamefont{A.}~\bibnamefont{Arima}} \bibnamefont{and}
  \bibinfo{author}{\bibfnamefont{F.}~\bibnamefont{Iachello}},
  \bibinfo{journal}{Phys. Rev. C} \textbf{\bibinfo{volume}{16}},
  \bibinfo{pages}{2085} (\bibinfo{year}{1977}),
  \urlprefix\url{https://link.aps.org/doi/10.1103/PhysRevC.16.2085}.

\bibitem[{\citenamefont{Pascu et~al.}(2009)\citenamefont{Pascu, C\u{a}ta-Danil,
  Bucurescu, M\ifmmode~\u{a}\else \u{a}\fi{}rginean, Zamfir, Graw, Gollwitzer,
  Hofer, and Valnion}}]{pascu2009}
\bibinfo{author}{\bibfnamefont{S.}~\bibnamefont{Pascu}},
  \bibinfo{author}{\bibfnamefont{G.}~\bibnamefont{C\u{a}ta-Danil}},
  \bibinfo{author}{\bibfnamefont{D.}~\bibnamefont{Bucurescu}},
  \bibinfo{author}{\bibfnamefont{N.}~\bibnamefont{M\ifmmode~\u{a}\else
  \u{a}\fi{}rginean}}, \bibinfo{author}{\bibfnamefont{N.~V.}
  \bibnamefont{Zamfir}},
  \bibinfo{author}{\bibfnamefont{G.}~\bibnamefont{Graw}},
  \bibinfo{author}{\bibfnamefont{A.}~\bibnamefont{Gollwitzer}},
  \bibinfo{author}{\bibfnamefont{D.}~\bibnamefont{Hofer}}, \bibnamefont{and}
  \bibinfo{author}{\bibfnamefont{B.~D.} \bibnamefont{Valnion}},
  \bibinfo{journal}{Phys. Rev. C} \textbf{\bibinfo{volume}{79}},
  \bibinfo{pages}{064323} (\bibinfo{year}{2009}),
  \urlprefix\url{https://link.aps.org/doi/10.1103/PhysRevC.79.064323}.

\bibitem[{\citenamefont{Pascu et~al.}(2010)\citenamefont{Pascu, C\u{a}ta-Danil,
  Bucurescu, M\ifmmode~\u{a}\else \u{a}\fi{}rginean, M\"uller, Zamfir, Graw,
  Gollwitzer, Hofer, and Valnion}}]{pascu2010}
\bibinfo{author}{\bibfnamefont{S.}~\bibnamefont{Pascu}},
  \bibinfo{author}{\bibfnamefont{G.}~\bibnamefont{C\u{a}ta-Danil}},
  \bibinfo{author}{\bibfnamefont{D.}~\bibnamefont{Bucurescu}},
  \bibinfo{author}{\bibfnamefont{N.}~\bibnamefont{M\ifmmode~\u{a}\else
  \u{a}\fi{}rginean}},
  \bibinfo{author}{\bibfnamefont{C.}~\bibnamefont{M\"uller}},
  \bibinfo{author}{\bibfnamefont{N.~V.} \bibnamefont{Zamfir}},
  \bibinfo{author}{\bibfnamefont{G.}~\bibnamefont{Graw}},
  \bibinfo{author}{\bibfnamefont{A.}~\bibnamefont{Gollwitzer}},
  \bibinfo{author}{\bibfnamefont{D.}~\bibnamefont{Hofer}}, \bibnamefont{and}
  \bibinfo{author}{\bibfnamefont{B.~D.} \bibnamefont{Valnion}},
  \bibinfo{journal}{Phys. Rev. C} \textbf{\bibinfo{volume}{81}},
  \bibinfo{pages}{014304} (\bibinfo{year}{2010}),
  \urlprefix\url{https://link.aps.org/doi/10.1103/PhysRevC.81.014304}.

\bibitem[{\citenamefont{Kotila et~al.}(2012)\citenamefont{Kotila, Nomura, Guo,
  Shimizu, and Otsuka}}]{kotila2012}
\bibinfo{author}{\bibfnamefont{J.}~\bibnamefont{Kotila}},
  \bibinfo{author}{\bibfnamefont{K.}~\bibnamefont{Nomura}},
  \bibinfo{author}{\bibfnamefont{L.}~\bibnamefont{Guo}},
  \bibinfo{author}{\bibfnamefont{N.}~\bibnamefont{Shimizu}}, \bibnamefont{and}
  \bibinfo{author}{\bibfnamefont{T.}~\bibnamefont{Otsuka}},
  \bibinfo{journal}{Phys. Rev. C} \textbf{\bibinfo{volume}{85}},
  \bibinfo{pages}{054309} (\bibinfo{year}{2012}),
  \urlprefix\url{https://link.aps.org/doi/10.1103/PhysRevC.85.054309}.

\bibitem[{\citenamefont{{J. Bartel {\it et al.}}}(1982)}]{bartel1982}
\bibinfo{author}{\bibnamefont{{J. Bartel {\it et al.}}}},
  \bibinfo{journal}{Nucl. Phys. A} \textbf{\bibinfo{volume}{386}},
  \bibinfo{pages}{79} (\bibinfo{year}{1982}), ISSN \bibinfo{issn}{03759474}.

\bibitem[{\citenamefont{Nomura et~al.}(2010)\citenamefont{Nomura, Shimizu, and
  Otsuka}}]{nomura2010}
\bibinfo{author}{\bibfnamefont{K.}~\bibnamefont{Nomura}},
  \bibinfo{author}{\bibfnamefont{N.}~\bibnamefont{Shimizu}}, \bibnamefont{and}
  \bibinfo{author}{\bibfnamefont{T.}~\bibnamefont{Otsuka}},
  \bibinfo{journal}{Phys. Rev. C} \textbf{\bibinfo{volume}{81}},
  \bibinfo{pages}{044307} (\bibinfo{year}{2010}).

\bibitem[{\citenamefont{Bonche et~al.}(2005)\citenamefont{Bonche, Flocard, and
  Heenen}}]{ev8}
\bibinfo{author}{\bibfnamefont{P.}~\bibnamefont{Bonche}},
  \bibinfo{author}{\bibfnamefont{H.}~\bibnamefont{Flocard}}, \bibnamefont{and}
  \bibinfo{author}{\bibfnamefont{P.~H.} \bibnamefont{Heenen}},
  \bibinfo{journal}{Compt. Phys. Commun.} \textbf{\bibinfo{volume}{171}},
  \bibinfo{pages}{49} (\bibinfo{year}{2005}).

\bibitem[{\citenamefont{Nomura}(2012)}]{nomura2012phd}
\bibinfo{author}{\bibfnamefont{K.}~\bibnamefont{Nomura}}, Ph.D. thesis,
  \bibinfo{school}{The University of Tokyo} (\bibinfo{year}{2012}).

\bibitem[{\citenamefont{Scholten}(1980)}]{scholten1980phd}
\bibinfo{author}{\bibfnamefont{O.}~\bibnamefont{Scholten}}, Ph.D. thesis,
  \bibinfo{school}{University of Groningen} (\bibinfo{year}{1980}).

\bibitem[{\citenamefont{Warner and Casten}(1983)}]{warner1983}
\bibinfo{author}{\bibfnamefont{D.~D.} \bibnamefont{Warner}} \bibnamefont{and}
  \bibinfo{author}{\bibfnamefont{R.~F.} \bibnamefont{Casten}},
  \bibinfo{journal}{Phys. Rev. C} \textbf{\bibinfo{volume}{28}},
  \bibinfo{pages}{1798} (\bibinfo{year}{1983}),
  \urlprefix\url{https://link.aps.org/doi/10.1103/PhysRevC.28.1798}.

\bibitem[{\citenamefont{McCutchan et~al.}(2004)\citenamefont{McCutchan, Zamfir,
  and Casten}}]{maccutchan2004}
\bibinfo{author}{\bibfnamefont{E.~A.} \bibnamefont{McCutchan}},
  \bibinfo{author}{\bibfnamefont{N.~V.} \bibnamefont{Zamfir}},
  \bibnamefont{and} \bibinfo{author}{\bibfnamefont{R.~F.}
  \bibnamefont{Casten}}, \bibinfo{journal}{Phys. Rev. C}
  \textbf{\bibinfo{volume}{69}}, \bibinfo{pages}{064306}
  (\bibinfo{year}{2004}),
  \urlprefix\url{https://link.aps.org/doi/10.1103/PhysRevC.69.064306}.

\bibitem[{\citenamefont{{Brookhaven National Nuclear Data Center}}()}]{data}
\bibinfo{author}{\bibnamefont{{Brookhaven National Nuclear Data Center}}},
  \bibinfo{howpublished}{{http://www.nndc.bnl.gov}}.

\bibitem[{\citenamefont{Casten and Zamfir}(2001)}]{casten2001}
\bibinfo{author}{\bibfnamefont{R.~F.} \bibnamefont{Casten}} \bibnamefont{and}
  \bibinfo{author}{\bibfnamefont{N.~V.} \bibnamefont{Zamfir}},
  \bibinfo{journal}{Phys. Rev. Lett.} \textbf{\bibinfo{volume}{87}},
  \bibinfo{pages}{052503} (\bibinfo{year}{2001}).

\bibitem[{\citenamefont{Debenham and Hintz}(1972)}]{DEBENHAM1972385}
\bibinfo{author}{\bibfnamefont{P.}~\bibnamefont{Debenham}} \bibnamefont{and}
  \bibinfo{author}{\bibfnamefont{N.~M.} \bibnamefont{Hintz}},
  \bibinfo{journal}{Nuclear Physics A} \textbf{\bibinfo{volume}{195}},
  \bibinfo{pages}{385 } (\bibinfo{year}{1972}), ISSN \bibinfo{issn}{0375-9474},
  \urlprefix\url{http://www.sciencedirect.com/science/article/pii/0375947472910676}.

\bibitem[{\citenamefont{Burke et~al.}(1988)\citenamefont{Burke,
  L\o{}vh\o{}iden, and Thorsteinsen}}]{burke1988}
\bibinfo{author}{\bibfnamefont{D.}~\bibnamefont{Burke}},
  \bibinfo{author}{\bibfnamefont{G.}~\bibnamefont{L\o{}vh\o{}iden}},
  \bibnamefont{and}
  \bibinfo{author}{\bibfnamefont{T.}~\bibnamefont{Thorsteinsen}},
  \bibinfo{journal}{Nuclear Physics A} \textbf{\bibinfo{volume}{483}},
  \bibinfo{pages}{221 } (\bibinfo{year}{1988}), ISSN \bibinfo{issn}{0375-9474},
  \urlprefix\url{http://www.sciencedirect.com/science/article/pii/0375947488905337}.

\bibitem[{\citenamefont{Kolata and Oothoudt}(1977)}]{kolata1977}
\bibinfo{author}{\bibfnamefont{J.~J.} \bibnamefont{Kolata}} \bibnamefont{and}
  \bibinfo{author}{\bibfnamefont{M.}~\bibnamefont{Oothoudt}},
  \bibinfo{journal}{Phys. Rev. C} \textbf{\bibinfo{volume}{15}},
  \bibinfo{pages}{1947} (\bibinfo{year}{1977}),
  \urlprefix\url{https://link.aps.org/doi/10.1103/PhysRevC.15.1947}.

\bibitem[{\citenamefont{Maher et~al.}(1972)\citenamefont{Maher, Kolata, and
  Miller}}]{maher1972}
\bibinfo{author}{\bibfnamefont{J.~V.} \bibnamefont{Maher}},
  \bibinfo{author}{\bibfnamefont{J.~J.} \bibnamefont{Kolata}},
  \bibnamefont{and} \bibinfo{author}{\bibfnamefont{R.~W.}
  \bibnamefont{Miller}}, \bibinfo{journal}{Phys. Rev. C}
  \textbf{\bibinfo{volume}{6}}, \bibinfo{pages}{358} (\bibinfo{year}{1972}),
  \urlprefix\url{https://link.aps.org/doi/10.1103/PhysRevC.6.358}.

\end{thebibliography}

\end{document}